\documentstyle[aas2pp4]{article}
\begin{document}

\title{Molecular Evolution in Collapsing Prestellar Cores}
\author{\sc Yuri Aikawa}
\affil{Department of Earth and Planetary Sciences, Faculty of Science,
Kobe University,\\ Kobe 657-8501, Japan}
\author{\sc Nagayoshi Ohashi}
\affil{Academia Sinica Institute of Astronomy and Astrophysics,\\
P.O. Box 1-87, Nankang, Taipei 115, Taiwan}
\author{\sc Shu-ichiro Inutsuka}
\affil{National Astronomical Observatory of Japan,\\
Mitaka, Tokyo 181-8588, Japan}
\author{\sc Eric Herbst}
\affil{Departments of Physics and Astronomy, The Ohio State
University, \\ Columbus, OH 43210}
\author{\sc Shigehisa Takakuwa}
\affil{Academia Sinica Institute of Astronomy and Astrophysics,\\
P.O. Box 1-87, Nankang, Taipei 115, Taiwan}

\begin{abstract}
We have investigated the evolution and distribution of 
molecules in collapsing
prestellar cores via numerical chemical models, adopting the Larson-Penston
solution and its delayed analogues to study collapse.
Molecular abundances and distributions in a collapsing core are
determined by the balance among the dynamical,
chemical and adsorption time scales.
When the central density $n_{\rm H}$ of a prestellar core
with the Larson-Penston flow
rises to $3\times 10^6$ cm$^{-3}$, the
CCS and CO column densities are calculated to show  central
holes of radius
7000 AU and 4000 AU, respectively,
while the column density of N$_2$H$^+$ is centrally peaked. These predictions
are consistent with observations of L1544.
If the dynamical time scale of the core is larger than that of
the Larson-Penston solution owing to magnetic fields, rotation,
or turbulence, the column densities of CO and CCS are smaller, and
their holes are larger than in the Larson-Penston core with the same central
gas density. On the other hand, N$_2$H$^+$ and NH$_3$ are more abundant
in the more slowly collapsing core. Therefore, molecular distributions
can probe the collapse time scale of
prestellar cores.
Deuterium fractionation has also been studied via numerical
calculations.
The deuterium fraction in molecules increases as a core evolves and
molecular depletion onto grains proceeds. When the central density
of the core is
$n_{\rm H}= 3\times 10^6$ cm$^{-3}$, the ratio DCO$^+$/HCO$^+$ at the
center is in the range 0.06-0.27, depending on the collapse time scale and
adsorption energy; this range is in reasonable agreement with the observed
value in L1544.

\end{abstract}


\section{INTRODUCTION}
Stars are formed in dense molecular cloud cores. Observations of cloud cores
are very useful in the study of star formation; the structures
and kinetics of star forming cores can be estimated via continuum
and molecular line observations (Myers 1985; Benson \& Myers 1989;
Williams et al. 2000). Many cores with a young embedded
protostar have been found and investigated following the
infrared survey by IRAS
(Beichman et al. 1986). These observations tell us about protostars and
their envelopes {\em after} the formation of a protostar.
In recent years, on the other hand, much effort has been done in the
observation of cores in which the protostar is not yet formed,
in order to understand the initial conditions and the earliest stages of
star formation (Andr$\acute{\rm e}$ et al. 2000).
Such dense molecular cores with
no protostars are called starless or prestellar cores.

L1544 is one such prestellar core, and it is the most extensively studied
so far. No sign of a protostar can be found toward L1544 by infrared
and radio cm continuum observations. Detailed observations of mm dust
continuum emission show that the gas density is high but almost constant,
with the number density of hydrogen molecules $n$(H$_2$)$=1.5\times 10^6$
cm$^{-3}$
inside a distance of 3000 AU from the emission peak
(Ward-Thompson, Motte, \& Andr$\acute{\rm e}$ 1999), also
suggesting that the central star is not yet formed.
In addition, molecular lines show a Doppler shift due to inward
motion on the order of 0.1 km s$^{-1}$ (Tafalla et al. 1998;
Williams et al. 1999; Ohashi et al. 1999).
Thus, L1544 must be in a very early stage of star formation.

Another interesting feature of L1544 is that there is a significant
chemical differentiation within the core. Figure \ref{fig:obs_dist}
shows that CCS is significantly depleted at the center
of the core (Ohashi et al. 1999), while N$_2$H$^+$ is centrally peaked
near the dust peak (Tafalla et al. 1998; Williams et al. 1999).
  The molecule
C$^{17}$O shows a similar distribution to that of CCS, with a central hole
of radius 6500 AU (Caselli et al. 1999).
Such chemical differentiation has also been found in the prestellar cores
L1498 (Kuiper, Langer \& Velusamy 1996) and L1521F (Ohashi 2000 and references
therein).
In general, molecular abundances are essential 
in determining
the structure of a core from line observations.
Thus, it is important to investigate
molecular distributions in collapsing cores with theoretical models.

In addition, the chemical differentiation within L1544 and similar
objects
can be a probe to investigate the infall time scale of the core, which
is an important parameter for understanding the physics of star
formation.  Molecular distributions are determined by the balance
between dynamical and chemical time scales.  For example, molecular
abundances should be constant throughout the core if the infall is
much more rapid than the molecular evolution.  If star formation
is purely determined by gravity, the core should evolve roughly in a
free-fall time scale.  If other processes, such as magnetic field,
rotation, or turbulence, are working against the collapse, the infall
velocity is smaller, and the infall time scale should be much larger
than the free-fall time scale.


Rawlings et al. (1992) have examined the evolution and
distribution of molecular
abundances in a collapsing core with the inside-out collapse solution of
Shu (1977).
This model corresponds to
a protostellar envelope in which a central object exists, because the
initial condition for inside-out collapse is a singularity at the center.
Their results cannot be applied to prestellar cores such as L1544,
in which the gas density is almost constant at the center. Bergin \& Langer
(1997) (See also Bergin 2000), on the other hand, have investigated molecular
evolution in a prestellar core extensively. Two different dynamical models
were examined: a dense cloud core contracting via ambipolar diffusion
(Basu \& Mouschovias 1994), and a phenomenological model where a static
core is accreting material from a low density halo.
They showed clearly that, in the collapsing core, sulphur-bearing molecules
such as CCS deplete earlier than nitrogen-bearing molecules such as
N$_2$H$^+$, which is qualitatively consistent with the observation of L1544.
They also showed that molecular evolution in each of their
two collapse models shows a similar behavior as a function of time, and
concluded that it is difficult to use chemistry to discriminate between
different dynamical solutions. However, they investigated molecular
evolution only with a single fluid element, and thus could not
predict detailed
distributions of molecular abundances, which should be compared with
observations. Molecular distributions could be a probe to
discriminate among dynamical solutions of the core.

This paper presents a theoretical model of the evolution of molecular
{\em distributions} in {\em prestellar} cores, and constrains the
collapse
time scale of the core by comparison of the theoretical results with the
observational data of L1544.
As a standard model of the collapsing core, we adopt the so-called
Larson-Penston collapse, which is a self-similar solution for a purely
dynamical isothermal collapse with spherical symmetry (Larson 1969;
Penston 1969). The solution yields a density
distribution in the core as a function of time in a semi-analytical
form, which enables us to calculate molecular distributions
in some evolutionary stages easily. In contrast to the inside-out
collapse, the Larson-Penston solution corresponds to collapse prior to
the formation of a protostar (Whitworth \& Summers 1985), and thus is suitable
for the study
of prestellar cores. Numerical simulations show that the inner region
of this solution is a good approximation for the ``first collapse''
phase of protostar formation (Masunaga et al. 1998; Masunaga \& Inutsuka 2000).

The remainder of the paper is organized as follows. The dynamical model of
the core and the model of molecular evolution are described in \S 2.
\S 3 is concerned with calculated molecular distributions 
in a core undergoing Larson-Penston
collapse, and the dependence of these molecular distributions
on the collapse time scale, the sticking probability of gaseous molecules
onto grain surfaces, the initial conditions, and adsorption
energies for gas-phase species onto grains.
In \S 4, the theoretical results are compared with the observations of L1544.
Deuterium fractionation in the collapsing core is also discussed.
Our summary is contained in \S 5.

\section{Model}
\subsection{Model of Collapsing Core}
We adopt the Larson-Penston solution (Larson 1969; Penston 1969) as our
standard model
for collapsing prestellar cores. The initial central density of the
cloud is assumed to be $n_{\rm H}=2.0\times 10^4$ cm$^{-3}$, which is a
representative value for molecular clouds.
The temperature is assumed to be 10 K.
Figure \ref{fig:euler} shows the distribution of density $n_{\rm H}$ and
infall velocity in the core at some evolutionary stages.
The density distribution at $t=1.89\times 10^5$ yr
is similar to that in L1544; in the central region the density is almost
constant with $n_{\rm H}\approx 3\times 10^6$ cm$^{-3}$,
and in the outer region
the density decreases as $n_{\rm H}\propto R^{-2}$, in which $R$ is the
distance from the core center.
The central densities at $t=1.52\times 10^5$ yr and $2.00\times 10^5$ yr,
respectively, are smaller and larger than at $t=1.89\times 10^5$ yr
by an order of magnitude.
The density distribution ultimately approaches
$n_{\rm H}\propto R^{-2}$
throughout the core at $t= 2.06 \times 10^5$yr.

The numerical calculation is based on Lagrangian coordinates and follows
molecular evolution in infalling fluid elements (or shells). Using the
distribution of density and velocity at each time step (Figure
\ref{fig:euler}), a temporal variation of the gas density in each fluid
element is obtained. For instance, Figure \ref{fig:lagrange} shows the density
variation in several fluid elements as a function of time.
The initial radius
of the core is assumed to be $3.2\times 10^4$ AU, so that the core radius
at $t=1.89\times 10^5$ yr is about the same as that of L1544
($1.5\times 10^4$ AU).

\subsection{Model of Molecular Evolution}
The basic equations for molecular evolution are given by
\begin{eqnarray}
\frac{dx(i)}{dt}&=&\sum_{j} \alpha_{ij}x(j)
+\sum_{j,k} \beta_{ijk}x(j)x(k)n_{\rm H},
\label{eq:reaction}\\
x(i)&=&n(i)/n_{\rm H}
\end{eqnarray}
where $n(i)$ is the number density of species $i$,
and $\alpha_{ij}$ and $\beta_{ijk}$ are the rate coefficients.  The
first term on the right-hand side of equation (\ref{eq:reaction})
represents reactions with external particles such as ionization by
cosmic rays.  The second term represents two-body reactions in which
species $i$ are formed by the reactions of species $j$ and $k$, or
species $i$ ($=j$) are destroyed by reactions with $k$.

For elemental abundances, we assume the so-called ``low-metal'' values
(e.g.  Lee et al.  1998; Aikawa et al.  1999).  The initial molecular
abundances are listed in Table \ref{tab:initial}.
Gas phase species are in atomic or
ionized form except for hydrogen, which is in molecular (H$_2$) form.

We have adopted the so-called ``new standard model'' network of
chemical reactions for the gas-phase chemistry (Terzieva \& Herbst
1998; Aikawa \& Herbst 1999). The ionization rate by cosmic rays is assumed to
be $\zeta=1.3\times 10^{-17}$ s$^{-1}$.
In addition to reactions in the gas phase,
the formation of H$_2$ molecules and the recombination of ions and electrons
on grain surfaces are taken into account.
The formation of ice mantles due to adsorption of
gaseous molecules and the desorption of molecules from ice mantles are
explicitly followed. The sticking probability $S$ of neutral species
when they collide with a grain is assumed to be 1.0 (Williams 1993).
Uncertainty in the sticking probability and its effect on our results will be
discussed later (\S 3.4).
We adopt a grain radius of $1.0\times 10^{-5}$ cm, which is considered to
be a typical value in interstellar clouds.
Two kinds of desorptive mechanisms are included:
thermal desorption, and non-thermal
desorption via the impulsive heating of cosmic rays (L$\acute{\rm e}$ger
et al. 1985; Hasegawa \& Herbst 1993). For simplicity and clarity,
other chemical reactions on grain surfaces are not considered because the
the rates of grain surface reaction are still controversial
(Caselli, Hasegawa,
\& Herbst 1998; Shalabiea, Caselli, \& Herbst 1998).

In the outer region of the core, photodissociation dramatically reduces
molecular abundances if the core is directly exposed to interstellar
ultra-violet radiation. Considering that CCS is detected even in the
outermost region of the core ($R\approx15000$ AU), and that the core is
embedded in a molecular cloud, $A_{\rm v}=3$ mag is assumed
at the outer boundary
of the core, so that the photodissociation does not much affect our model.

\section{RESULTS}
\subsection{Molecular Evolution in an Infalling Fluid Element}
Figure \ref{fig:evol} shows the evolution of CO, CCS, N$_2$H$^+$ and some
other species in an infalling fluid element.
Molecules adsorbed onto grain
surfaces are referred to as ``ice'' in the Figure and hereinafter.
Gas-phase molecules decrease sharply in abundance
at $t\sim 2\times10^5$ yr, mainly by adsorption onto grains.
For the three gas-phase species we are mainly concerned with
-- CO , CCS, and N$_2$H$^+$ -- the depletion time scales differ slightly.
The abundance of the radical CCS decreases
first, due both to adsorption
and chemical reactions: CCS reacts with H$_3$O$^+$ or
HCO$^+$ to produce HC$_2$S$^+$, which is partially transformed to CS and
then to CO. Even in a purely gas-phase model of molecular clouds with
constant density ($n_{\rm H}\sim 10^4$ cm$^{-3}$), CCS is well-known
as an ``early-time species'', the abundance of which has a peak at
$t\sim 10^5$ yr, and decreases afterwards.
The depletion of CO is mostly due to adsorption. The depletion
time scale of N$_2$H$^+$ is larger than for the other species
for the following three reasons. Firstly, N$_2$H$^+$ is not directly
adsorbed onto
grains. Since most grains are negatively charged at these densities
(Umebayashi \& Nakano 1980; Nakano \& Umebayashi 1986), N$_2$H$^+$
recombines to produce N$_2$, which returns to the gas phase and produces
N$_2$H$^+$ again. Secondly, the nitrogen molecule N$_2$,
which is the precursor of N$_2$H$^+$,
is produced in the late stages of the gas-phase chemistry,
and has a small binding
energy on a grain surface. Thirdly, the main destroyers of N$_2$H$^+$
in the gas phase,
CO and electrons, decrease in the later stages due to adsorption and the
increased gas density. The N$_2$H$^+$ ion decreases in abundance eventually
owing to the depletion of the main precursor, N$_2$
(Bergin \& Langer 1997).

\subsection{Spatial Distribution of Molecules}
Radial distributions of molecules within
the collapsing core at some fixed time steps are obtained by
performing molecular evolution calculations similar to that described
in the previous subsection for many fluid elements in the flow.
Figure \ref{fig:dist} shows such distributions in terms of absolute
abundances (cm$^{-3}$).

At $t=1.52\times 10^5$ yr, most species in the figure are slightly more
abundant in the inner region of the core, while CCS and C$_3$H$_2$ are
less abundant in this region, where the
  gas density $n_{\rm H}$
is higher. If we plotted molecular abundances
relative to hydrogen nuclei, instead of plotting absolute abundances,
only SO$_2$ and N$_2$H$^+$ would be more abundant
in the inner regions.

As the core evolves, the gas density in the central region becomes higher,
which accelerates a further depletion of gaseous molecules.
At $t=1.89\times 10^5$ yr, the spatial variation of the molecular density
can be more pronounced than at $t=1.52\times 10^5$ yr. The molecules
CCS and CO are depleted within a radius of several thousand AU,
while N$_2$H$^+$ and NH$_3$ are still more abundant in the inner regions.
At $t=2.00\times 10^5$ yr, even these nitrogen-containing molecules are
slightly depleted at $R < $ 1000 AU.

In order to compare our results with the observational data, molecular
column densities in a spherical core are obtained by integrating along
the line of sight. The effect of the integration is to average out the
density dependence on $R$  along the line of sight while preserving it
in a direction perpendicular to the observer.  Although the column density
distribution is not as sharp as the distribution of abundances,
it is unusual for the averaging to
eliminate important radial dependences.
Figure \ref{fig:column} shows distributions of column
densities for assorted species at the three stages.
Column density
distributions for a larger number of species at $t=1.89\times 10^5$ yr are
available in Table \ref{tab:elect}, designed for electronic presentation.
CCS and C$_3$H$_2$ have a central hole
at these three stages. On the other hand N$_2$H$^+$ and NH$_3$ are centrally
peaked most of the time.
At $t=1.52\times 10^5$ yr CO and HCO$^+$ are centrally peaked, and 
the CS column
density is almost constant at the center, while at later stages these species
have a central hole. Column densities
of some species at the center do not monotonically change as the
collapse proceeds; e.g., the central
SO$_2$ column density increases from $t=1.52\times 10^5$ yr
to $1.89\times 10^5$yr, but then decreases afterwards.

At $t=1.89 \times 10^5$ yr, both the gas density
and the molecular distributions are consistent with the observed
characteristics of L1544. The dust continuum shows a gas density of
$n_{\rm H}\sim 3 \times 10^6$ cm$^{-3}$ within the central region of radius
about 3000 AU, while molecular observations show that CCS and CO have a central
hole of radius about 7000 AU (Caselli et al. 1999; Ohashi et al. 1999),
and N$_2$H$^+$ has a peak near the dust peak (Tafalla et al. 1998;
Williams et al. 1998).
More quantitative comparisons are discussed in \S 4.1.

\subsection{Dependence on Collapse Time Scale}
If a core is partially supported
by rotation (Matsumoto, Hanawa, \& Nakamura 1997; Saigo \& Hanawa 1998),
magnetic fields (Basu \& Mouschovias 1994) or turbulence,
the collapse proceeds more slowly than in the Larson-Penston
solution. This subsection shows
how the molecular distribution depends on the collapse time scale in order to
determine whether the dynamics of the collapse can be 
distinguished from molecular
observations. It would be preferable if we could adopt an analytical
or semi-analytical solution for the case of slow collapse due to
magnetic fields or turbulence.
Because no such solution is available, slow collapse models
are constructed
by artificially slowing down
the Larson-Penston collapse by a constant factor;
the inward velocity of the gas is divided by a factor $f$ of 3 and 10.

Figure \ref{fig:evol_slow} shows molecular evolution in fluid elements
for the Larson-Penston collapse and for a slower collapse with $f=10$.
Figure \ref{fig:evol_slow_R} shows results of the same evolution calculation,
but as a function
of the gas density in the fluid element at each moment.
The molecular evolution in later stages
($t\gtrsim 10^5$ yr) can be seen more clearly in Figure \ref{fig:evol_slow_R}.
As discussed previously for the $f=1$ case, the fractional
abundances
in Figure \ref{fig:evol_slow} (and Figure \ref{fig:evol_slow_R})
are determined by a competition among chemical,
adsorption and dynamical time scales. This competition is complex and is
different for different species. For a molecule made relatively early
and which is depleted mainly by adsorption onto grains at high densities,
a slower collapse allows adsorption to proceed more efficiently. Carbon
monoxide falls in this category, and the abundance of this species at
$R=3000$ AU (the inner radius considered) in Figure \ref{fig:evol_slow}
is smaller in (b) than in (a). The abundance of CCS, an early-time species,
at $R=3000$ AU is also significantly smaller in (b) than (a).
On the other hand, N$_2$H$^+$ and NH$_3$ increase
significantly at late stages in the slow collapse (b), when compared with
(a). These so-called ``late-time'' species are produced slowly
in the gas. In addition, depletion of the main destroying reactant
--- CO for N$_2$H$^+$, and HCO$^+$ for NH$_3$--- enhances their abundance
(Bergin \& Langer 1997). While SO$_2$ increases to $n({\rm SO}_2)/n_{\rm H}
\sim 10^{-10}$ within $2\times 10^5$ yr in Figure \ref{fig:evol_slow} (a),
it is almost constant ($\sim 10^{-12}$) at $t\sim 10^5$ yr, and increases
to $\sim 10^{-10}$ at $t\sim 1\times 10^6$ yr in the slow collapse (b),
because its main formation path SO + O is more efficient at higher density.
A decrease of HCO$^+$, which is a principal destroyer of SO, also enhances
the abundance of SO$_2$ (Bergin \& Langer 1997).
These formation reactions all become less efficient than adsorption
in the final moment of the studied collapse, and the abundance of SO$_2$
at  $R=3000$ AU is significantly smaller in the slow collapse (b) than in the
Larson-Penston case (a).

Figure \ref{fig:dist_slow} shows distributions of assorted absolute molecular
abundances at the three evolutionary
stages of the core for collapse slowed by factors
of $f=3$ and $f=10$.
The distribution of the total (hydrogen) density in the core is 
exactly the
same as in Figure \ref{fig:dist} for each panel (a-c), so these cores will be
recognized as in the same physical evolutionary stage, if observed by dust
emission.
Comparison of Figures \ref{fig:dist_slow} and  \ref{fig:dist} shows
that the molecular distribution clearly depends on the collapse model.
The depletion of CO and CCS is more severe in the slow collapse cases,
especially in the inner core (ex. $R\lesssim 5000$ AU at $t=f \times 1.89
\times 10^5$ yr). The 
radius of the region in which the absolute
abundance of SO$_2$ has a peak value is larger in the slow collapse cases.
The late-time species --- NH$_3$, and N$_2$H$^+$ --- are more abundant in the
slow collapse cases than in the Larson-Penston case (Figure \ref{fig:dist}),
because of their slow formation in the gas phase and the depletion of some
species which destroy them.

The column densities of assorted species at $t=f\times 1.89\times 10^5$ yr
for all collapse models (f = 1, 3, and 10) are shown in Figure 
\ref{fig:column_slow}.
In spite of the fact that the physical structure of the core is
exactly the same, the distributions of molecular column
densities are significantly different from each other.
The distributions of CO, CCS, CS, C$_3$H$_2$ and SO$_2$ show a hole
structure, and the width of the hole is larger (if not deeper) in the
slower collapse cases.
The column densities of CO, CCS, CS and C$_3$H$_2$ are smaller in the
cases of slow collapse,
while the column density of SO$_2$ does not change monotonically
as a function of $f$.
On the other hand, N$_2$H$^+$ and NH$_3$ are centrally peaked in all three
models, and their column density is larger for slower collapse.
Clearly the molecular column densities and
distributions in prestellar cores are strongly dependent on the collapse
time scale, and can therefore probe the dynamics of collapsing cores,
more specifically if collapse occurs in a free-fall time scale or more slowly
owing to partial support by rotation, magnetic fields or turbulence.

\subsection{Dependence on the Sticking Probability}
The previous subsection showed that adsorption of gaseous
molecules onto grains is a major contributor to molecular distributions
in cores. The sticking probability $S$ of molecules when they collide
with a grain is, however, rather uncertain. Based on some experimental
and theoretical work, $S$ is estimated to be $0.1-1.0$ at temperatures
of $\sim 10$ K (Williams 1993).
So far $S=1.0$ has been assumed.

The dot-dashed lines in Figure \ref{fig:column_slow} show the column densities
of assorted molecules for the case with $S=0.33$ and $f=3$, while the solid
lines show the case with Larson-Penston collapse and $S=1.0$.
Since the time scale
of adsorption is inversely proportional to $S$, these two lines should
be close, if the molecular distribution is determined simply by the balance
between collapse and adsorption. The species CO and HCO$^+$ show such
behavior, which suggests that the collapse time scale cannot be estimated
based only on CO and HCO$^+$, unless the exact value of the sticking
probability is known.
On the other hand, the column densities of CCS and C$_3$H$_2$ are
much smaller for $S=0.33$ and $f=3$ than for $S=1.0$ and $f=1$,
because chemical reactions, as well as adsorption, are the main
destruction mechanism of these ``early-time species'', which
suggests that they can be good probes of the collapse time scale
and are relatively insensitive to $S$ (compare the dotted and
dot-dashed lines in Figure \ref{fig:column_slow}).
The column densities of SO$_2$, N$_2$H$^+$, and NH$_3$ are larger for
$S=0.33$ and $f=3$ than for $S=1$ and $f=1$,
because they are efficiently produced in the gas phase at late times.
Thus, a low column density of CCS reinforced by high column densities
of N$_2$H$^+$ and NH$_3$ can be considered as a characteristic of slow
collapse ($f > 1$), regardless of the uncertainty in the sticking probability.

\subsection{Dependence on the Initial Conditions}
The initial conditions pose another source of uncertainty in the molecular
evolution of a collapsing core. In the above calculations, the initial gas
was assumed to consist of atoms and ions, with total gas density
$n_{\rm H}\sim 2\times10^4$ cm$^{-3}$.

As an alternative, consider a Larson-Penston collapse 
starting from a gas density of
$n_{\rm H}(R=0)= 2.0\times10^2$ cm$^{-3}$, which is a representative value
for diffuse clouds. The initial gas consists of atoms and ions except for
hydrogen, which is in molecular form.
In this model, the density distribution reaches the same values
as that of the initial condition of the original model -- 
$n_{\rm}(R=0)=2.0\times
10^4$cm$^{-3}$ -- after $1.85\times 10^6$ yr. Then the subsequent evolution
of the density distribution is the same as that of the original model with
the time $t$ of the calculation shifted by $1.85\times 10^6$ yr;
the central density
of $n_{\rm H}= 3.0\times 10^6$ cm$^{-3}$ is reached at $t=2.04\times 10^6$
yr, and the density distribution ultimately approaches $n_{\rm H}
\propto R^{-2}$ at $t=2.06\times 10^6$ yr. The dotted lines in Figure
\ref{fig:column_init} show molecular column densities for this model
at $t=2.04\times 10^6$ yr while the solid lines represent the original model
at $t=1.89\times 10^5$ yr. Although the temporal gas density ($n_{\rm H}$)
distribution is the same for these two cores, the column densities of the
assorted molecules are different. In the model with the lower
initial density, CCS, CO, CS, and C$_3$H$_2$ are less abundant,
and N$_2$H$^+$, NH$_3$, and SO$_2$ are more abundant, because of the
extended exposure to the chemistry, just as in the slow collapse model.
But the absolute differences are not significant --- only a
factor of 2 or so ---
because the gas spent the initial phase ($1.85\times
10^6$ yr) at low density. More importantly, the radial profiles
stay the same. Hence, our results are not
significantly dependent on the choice of the initial density of
the Larson-Penston collapse.

In a second alternative model, the physical evolution of
the collapse is the same as that of the original model, but the
gas consists mostly of molecules when the collapse starts.
During the first $t=3\times 10^5$ yr, the cloud is supported by turbulence
or magnetic fields, and the gas density stays constant with a typical molecular
cloud value of $n_{\rm H}=2.0\times 10^4$ cm$^{-3}$.
The duration of this pre-collapse phase is chosen because
the molecular evolutionary model at this constant density reproduces
the averaged abundances in molecular clouds, such as TMC-1 at around this
time very well (Millar, Farquhart \& Willacy 1997; Terzieva \& Herbst 1998).
Then
the support decays abruptly,
for simplicity, and the gas collapses with a Larson-Penston flow.
The dashed lines in Figure \ref{fig:column_init} show calculated molecular
column densities vs $R$ when the central density of the core reaches
$n_{\rm H}=3.0\times 10^6$ cm$^{-3}$, i.e. $1.89\times 10^5$ yr after
the collapse starts.
The deviations from the original model are more significant than in
the first alternative model; CCS, CS, and C$_3$H$_2$ are 
decreased more than
an order of magnitude,
and SO$_2$ is increased by a factor of 5. Therefore, the duration of
the pre-collapse phase can have an important effect in determining
the molecular evolution in the collapsing core, if the gas density in the
pre-collapse phase is as high as $n_{\rm H}\sim 10^4$ cm$^{-3}$.
The deviation is less significant if the duration of the pre-collapse
is shorter, or the density of the pre-collapse stage is lower.

\subsection{Uncertainty in the Adsorption Energy}
The desorption rate of molecules from the grain surface is another
uncertainty in the theoretical model. Both thermal and non-thermal
desorption rates are sensitively dependent on the adsorption (binding) energy
of a molecule to the grain surface, but experimental data of
adsorption energies are not yet available for many of the species
included in our model. In addition, the binding energy is
dependent on the nature of the grain surface, about which little 
is known,
and which is likely to be time-dependent.
For example, the binding energy of a CO molecule is 960 K on pure CO ice,
and it is 1740 K on water ice (Sandford \& Allamandola 1988;
Sandford \& Allamandola 1990).
For simplicity and consistency with previous work, we have so far adopted
adsorption energies for an SiO$_2$ surface, obtained for the most part
theoretically by integrating the van der Waals interaction between a molecule
and the surface (Allen \& Robinson 1977; Hasegawa \& Herbst 1993).
This model is labeled as Model A hereinafter.

As an alternative (hereinafter Model B), some of the 
adsorption energies
are replaced by the experimental data on pure ices. The modified
values are listed in Table \ref{tab:adsorb} together with the original
values (Yamamoto, Nakagawa, \& Fukui 1983; Sandford \& Allamandola 1993;
Aikawa et al. 1997). In Model B, the adsorption energies of non-polar
molecules such as CO and N$_2$ are smaller, and the adsorption
energies of polar molecules such as H$_2$O larger than in Model A. For
species not listed in Table \ref{tab:adsorb}, the same values are adopted
as in Model A (Hasegawa \& Herbst 1993). Another alternative (Model C)
corresponds to the ``H$_2$O ice mantle'' of Bergin \&
Langer (1997), in which the original values are multiplied by a constant
factor 1.47, which is a ratio of the binding energy of CO on H$_2$O ice
to that on an SiO$_2$ surface.

Figure \ref{fig:column_Eads} shows assorted column densities at $t=1.89\times
10^5$ yr for the three models.
The Larson-Penston collapse is assumed.
The radial distribution of the column density
of CCS is not much affected by the varied assumptions concerning the
adsorption energy. On the other hand, the CO column density in Model B
does not show a central hole, which is apparent in Model A, because net
adsorption is too slow.
If the ice mantle of the grain is as non-polar as pure CO ice, a slower
collapse than the Larson-Penston collapse is preferable to account
for the CO observations in L1544. The species N$_2$H$^+$ and NH$_3$ are not
very sensitive to varying the
adsorption energy. Although the adsorption energy of NH$_3$
is much larger in Model B than in Model A, NH$_3$ is still very
abundant and centrally peaked because it is produced in the gas phase from
N$_2$ at later stages. At $t=2.00\times 10^5$ yr, however, N$_2$H$^+$
and NH$_3$ have a central hole of radius about 3000 AU and several hundred AU,
respectively, in Model C. The species SO$_2$ is rather sensitive to the
uncertainties of the adsorption energy; if the ice mantle is polar (Model C),
its column density is significantly lower, and shows a central
hole of radius
4000 AU, in contrast to the other two models. The ion HCO$^+$, which is
formed directly from
CO, is more abundant in the central region in Model B than in Model A.
In Model C, the column density of HCO$^+$ is higher than in Model A,
in spite of the lower column density of CO, because the main destroyers
of ions in the gas phase such as H$_2$O are less abundant in Model C.
At $R\lesssim 5000$ AU the column density of C$_3$H$_2$ is higher in Model
C than in other two models, because O atoms, which are a main destroyer of
carbon-chain species, are more efficiently depleted from the gas phase.
The depletion of O atoms in the inner regions has the same effect on the
abundances of CS and CCS, but there are not significant changes in their column
densities, which mainly derive from the outer regions.

\section{DISCUSSION}
\subsection{Comparison with L1544}
Our model results can be compared with observed molecular
abundances in the prestellar core L1544, in an attempt to determine
if the core is
collapsing in a free-fall time scale (i.e. Larson-Penston collapse), or
more slowly. The morphology of L1544 is in fact different from our
spherical model; it has a flattened structure.
But molecular column densities in L1544 can be compared with
those obtained in our model, because L1544 is in an almost edge-on
configuration (Ohashi et al. 1999), and because the free-fall time scale
does not significantly depend on the elongation of the core.

Table \ref{tab:comp} shows the characteristics of assorted
molecular distributions in L1544 and theoretical models with a central
gas density $n_{\rm H}=3\times 10^6$ cm$^{-3}$.
In the theoretical models the initial condition is atomic, and
Model A is assumed for the adsorption energies.
In L1544, CO and CCS show a central hole of radius $6-8$ thousand AU,
and the peak column densities of CCS and CO are  $4\times 10^{13}$
cm$^{-2}$ and $2\times10^{18}$ cm$^{-2}$, respectively
(Ohashi et al.1999; Caselli et al. 1999; Caselli 2000, private
communication).
The exact radius of the hole is difficult
to determine from the observational data because of the complex morphology
and clumpiness of the core, especially for the CCS line. Among the
four models listed in Table \ref{tab:comp}, the model with Larson-Penston
collapse (f=1, S=1.0) is the most consistent with the observed
distributions of CCS and CO, which means $f$ should not be much larger than 1.
In the Larson-Penston model the ratio of the CCS column density
at the peak compared with the core center is about 1.4, which is 
also in reasonable
agreement with the observed value.
One objection against the Larson-Penston model might be that the infall
velocity observed
in L1544 ($\sim 0.1$ km s$^{-1}$) is smaller than those in the outer region
of the model (Figure \ref{fig:euler}).
But this discrepancy is an effect of the outer boundary condition;
Larson-Penston flow is a self-similar solution and does not take into
account the fact that the flow velocity should be zero at a
certain radius.
In fact, numerical simulation of a collapsing core shows that
the velocity in the outer region of the core is significantly smaller
than the Larson-Penston solution, but that the dynamics of the core as a
whole are similar to that of the Larson-Penston solution
(Masunaga et al. 1998).

The column density of N$_2$H$^+$ is centrally peaked in L1544, and all
models listed in Table \ref{tab:comp} reproduce this distribution.
However, the absolute value of the N$_2$H$^+$ column density obtained
in our models is smaller than observed by a factor of $3-20$.
There are two possible explanations for this discrepancy.
Since the model column density is smaller than observed in all four cases, the
easiest solution would be to modify the reaction rate coefficient concerning
the formation of N$_2$H$^+$. The precursor molecule, N$_2$, is formed by
the reactions N + OH $\to$ NO + H and N + NO $\to$ N$_2$ + O.
Since the temperature dependence of the neutral-neutral reaction rates is not
well known, these reactions may be a source of error in the theoretical model.
But it is not  probable that the error in the reaction rate
coefficients causes the one-order-of-magnitude difference between the
observation and the Larson-Penston model.
Another explanation concerns the initial condition of the core formation.
So far we have assumed that all the gases in the core start chemical evolution
and collapse at the same time. But there is a possibility that the cloud
is initially more inhomogeneous, and the prestellar core is formed
by the coagulation of small clumps (Kuiper et al. 1996).
The results shown in the previous section and Table \ref{tab:comp}
indicate that the discrepancy can be resolved if the gas parcels traced
by N$_2$H$^+$ and CCS (and CO) have different ``chemical ages''.
While CCS cannot have
experienced chemical evolution for much more than the free-fall time scale,
the gas traced by N$_2$H$^+$ may have undergone chemical evolution for a
longer time scale,
because the N$_2$H$^+$ abundance increases as a function of time.
Such a time difference would be naturally produced in an inhomogeneous model,
with small clumps traced by CCS added to the central part of the core.
Low UV attenuation in the small clumps would help to keep their chemical age
young.
A central part with more clumps would form a large core with higher density
and attenuation by coagulation and collapse before the clumps in the
outer region accreted onto it, and thus molecular
evolution at the center would also proceed faster than in the outer region.
It is also consistent with the clumpy structure of the L1544 core observed
by the CCS line (Ohashi et al. 1999).

Readers may wonder if other uncertainties in the model, such as
additional desorption mechanisms (e.g. photodesorption in the surface region)
and temperature gradients in the core
could affect the molecular distribution and resolve the discrepancy
concerning CCS
and N$_2$H$^+$. For CCS, these uncertainties cannot change
our result significantly, since the CCS abundance decreases mainly because of
chemical reactions in the gas phase.
For N$_2$H$^+$, very little precursor
nitrogen resides on grain surfaces until very late times and the
chemistry of formation is not particularly temperature dependent.  So,
this ion is also not likely to be affected strongly by these
uncertainties.

Although we are confident of our qualitative conclusions, based
on various parameter studies in \S 3, it is also true that more work needs
to be done to constrain the uncertainties in chemical models, and to compare
the model results with the observational data quantitatively.
As one of such efforts, we are preparing another model calculation,
in which full grain-surface reactions are included.
More observations of different molecular species would be
helpful, as well, in order to clarify the uncertainties in chemical models,
the discrepancy concerning CCS and N$_2$H$^+$, and the physical 
evolution of prestellar
cores.

\subsection{Deuterium Fractionation}
Caselli et al. (1999) observed DCO$^+$ towards L1544, and found a high
abundance ratio for DCO$^+$/HCO$^+$ of $0.12\pm0.02$ and $0.04\pm 0.09$
at the position of the N$_2$H$^+$ emission peak
for two separate velocity components, corresponding to background (blue)
and foreground (red) components in the collapsing flow. In low
temperature
molecular clouds, ratios of deuterated to normal isotopes are much higher than
the cosmic elemental abundance D/H$\sim 1.5\times 10^{-5}$
because of the differences
in zero-point energies between deuterated and non-deuterated species,
and because of rapid isotopic exchange reactions (Millar, Bennet, \&
Herbst 1989). In addition, the deuterium isotopes are further enhanced
once the heavy molecules start to be depleted (Brown \& Millar 1989).
For example, H$_2$D$^+$ is mainly
formed by the reaction H$_3^+$ + HD $\to$ H$_2$D$^+$ + H$_2$ and destroyed
by the reactions with CO and with electrons. In most of the regions in our
calculation, CO is the main reactant; although the rate coefficient of the
reaction with electrons is about $3\times 10^2$ times larger than that with CO,
the electron abundance is much smaller than the CO abundance
(see Figures \ref{fig:dist} and \ref{fig:dist_slow}).
Thus the ratio H$_2$D$^+$/H$_3^+$ increases
as the CO is depleted. This larger ratio leads to enhanced
values for analogous ratios. In fact, the ratio DCO$^+$/HCO$^+ =0.12\pm 0.02$
in the blue-shifted component of L1544 is higher than the ratios
DCO$^+$/HCO$^+ =0.025-0.07$ in other
``average'' cloud cores in Taurus, which strengthens the argument that
the CO is depleted at the center of L1544 (Caselli et al. 1999).
Since DCO$^+$ and HCO$^+$ are formed by the reaction of H$_2$D$^+$ and
H$_3^+$ with CO, respectively, the ratio DCO$^+$/HCO$^+$ reflects
the ratio H$_2$D$^+$/H$_3^+$.
Deuterium chemistry has been included in our model
to see whether or not the observed deuterium fraction can be reproduced
and can yield some constraint on the model parameters.
A detailed description of our deuterium chemistry network
is found in Aikawa \& Herbst (1999).

Figure \ref{fig:D} (a) shows the calculated column density ratio of
DCO$^+$ to HCO$^+$ at three stages
for our standard model with $f=1$ and $S=1.0$.
The DCO$^+$/HCO$^+$ ratio increases as the core evolves,
because of the depletion of gas-phase species. Figure \ref{fig:D} (b)
compares our standard model with other cases.
The ratio is higher in all cases with a heavier depletion of CO. Considering
the rather wide range of the observed value in the ``red'' velocity
component of L1544 (DCO$^+$/HCO$^+=0.04\pm 0.09$), our models in Figure
\ref{fig:D} (b) are consistent with the observation, except for the slow
collapse with $f=10$, which yields much too high a result.
The value in the ``blue'' velocity component (DCO$^+$/HCO$^+ =0.12\pm
0.02$) is best reproduced in the case
of Larson-Penston collapse with a high adsorption energy (Model C), although
other models cannot be excluded because of the uncertainties in the
observational data and in the chemical reaction rate coefficients.

\section{SUMMARY}
The evolution and distribution of molecules in collapsing prestellar cores
have been investigated with the Larson-Penston collapse model.

The abundance for each molecule and
its distribution in a collapsing core
is determined by the balance between dynamical and chemical time scales
(including adsorption). Because of differing chemical
evolutionary patterns, molecular distributions differ for
different species. In our standard model, when the central density of the
core $n_{\rm H}$ rises to $3\times 10^6$ cm$^{-3}$ ($t=1.89\times 10^5$ yr),
the distributions of the CCS and CO column densities show a central hole
of radius 7000 AU and 4000 AU, respectively, while the column density of
N$_2$H$^+$ is centrally
peaked, all of which is consistent with observations of the prestellar core
L1544.
If the collapse time
scale is larger owing to rotation, magnetic fields or turbulence, the column
densities of CO and CCS are much smaller, and their central holes are larger
than in the case of Larson-Penston collapse. The species N$_2$H$^+$ and
NH$_3$, on the other hand, are more abundant with slower collapse, and
almost centrally peaked even if the collapse is slowed down by a factor of 10.

Several uncertainties in the theoretical model have been discussed,
including the
sticking probability onto grain surfaces, the initial physical and chemical
conditions, and the adsorption energies of molecules on grain surfaces.
Since the adsorption time scale is inversely proportional to the sticking
probability, the abundance of CO, which is mainly determined by adsorption,
is sensitively dependent on any assumption regarding the sticking probability.
On the other hand, the depletion time scale of CCS is less affected by the
sticking probability, because it is destroyed mainly by gas-phase chemistry.
Those species that show little dependence on $S$ are clearly better probes
of the dynamics.

A change of the initial density in our model does not significantly affect
the results;
even if the Larson-Penston collapse starts with a density $n_{\rm H}$ of
$2\times 10^2$ cm$^{-3}$, appropriate for diffuse clouds, molecular
column densities are different
from our standard model by at most a factor of 2 and the radial
distributions are similar when the central density of
the collapsing core reaches a certain value. The results are
affected if we include a pre-collapse stage lasting $3\times 10^5$ 
yr with gas
density $n_{\rm H}=2\times 10^4$ cm$^{-3}$.
Here the CCS column density is smaller than the standard case by more than
an order of magnitude when the central density of the core reaches
$3\times 10^6$ cm$^{-3}$, while the late-time species, especially SO$_2$
and N$_2$H$^+$,  are more abundant.
The difference from our standard model is less significant if the pre-collapse
stage is shorter or the gas density is smaller.

The dependence of our results on the uncertainty in grain
adsorption energy is mixed.
The CCS, N$_2$H$^+$, and NH$_3$ column
densities are not significantly affected by this uncertainty. The case
of CO is another story; if the grain surface is covered by non-polar
ice, and the adsorption
energy of CO is as low as that for pure CO ice, its column density does
not show the apparent hole structure at $t=1.89\times 10^5$ yr in the
Larson-Penston collapse, which is inconsistent with the observation of L1544.

Comparison of our results with the observed molecular column
densities
in L1544 shows that our standard model with Larson-Penston collapse
($S=1.0, f=1$) is more consistent with the distributions (column density and
hole radius) of CCS and CO than slower collapse models.
Since the column density of CCS is not much affected by the
uncertainty in sticking probability and adsorption energies, and is sensitively
dependent on the slowing down factor $f$, it gives a more
strict constraint
on the collapse time scale of L1544 than does CO.
Moreover, the calculated distributions of N$_2$H$^+$ in our models 
are all centrally
peaked, which is consistent with observation,
regardless of the collapse time scale. However, the absolute value of the
N$_2$H$^+$ column density is smaller than observed in all of our models,
and the agreement is the worst in the Larson-Penston case.
The most probable explanation is
an inhomogeneous initial condition of the core (Kuiper et al. 1996),
in which small
clumps traced by CCS coagulate 
with the central core with high density and UV attenuation traced
by N$_2$H$^+$.

The deuterium fraction in molecules increases as the core evolves and
heavy molecules reside on dust grains rather than in the gas.
When the central density of the core is $n_{\rm H}= 3\times10^6$ cm$^{-3}$,
the DCO$^+$/HCO$^+$ ratio at the center is calculated to lie
in the range of 0.06-0.27, depending on the collapse model and the
adsorption energies, which is in reasonable agreement with the observed
range in L1544.

In conclusion, we have shown that the distributions of 
molecular column densities
are probes of the collapse time scale of prestellar
cores, if compared with theoretical models quantitatively.
On the theoretical side, further consideration of uncertainties 
in the model,
such as the homogeneity of the initial conditions and grain-surface processes,
is desirable. On the observational side, mapping of other molecular lines of
both early-time species (e. g. carbon chains) and
late-time species (e. g. NH$_3$ and SO$_2$) would be helpful in
order to reveal
the evolution of the prestellar core L1544, and reduce
the uncertainties in the chemical models.
Statistical observations of molecular distributions in a large number of
prestellar cores are desirable in order to reach more general
conclusions on the physical and chemical evolution of these cores.

\acknowledgements
The authors are grateful to E. Bergin and D. Ruffle for useful comments.
Y. A. is grateful for financial support from the Japan Society for
Promotion of Science. N. O. is supported in part by NSC
Grant 89-2112-M-001-021. S. T. is supported by a Postdoctoral Fellowship
of Institute of Astronomy \& Astrophysics, Academia Sinica.
The Astrochemistry Program at The Ohio State
University is supported by The National Science Foundation.
Numerical calculations were carried out on the VX/4R at the Astronomical
Data Analysis Center of the National Astronomical Observatory
of Japan.



\clearpage

\figcaption[]{Total intensity map of N$_2$H$^+$ (contours,
Tafalla et al. 1998) and CCS (gray scale, Ohashi et al. 1999) in L1544.
The cross shows the peak of dust continuum emission.
\label{fig:obs_dist}}

\figcaption[]{Distribution of (a) density and (b) velocity in a collapsing
core with the  Larson-Penston solution.
The dotted lines show the initial condition ($t=0$) of the core.
The age of the core for other lines is $1.52\times10^5$ yr
({\em dashed lines}), $1.89\times 10^5$ yr ({\em solid lines}) and $2.00
\times10^5$ yr ({\em dot-dashed lines}).
\label{fig:euler}}

\figcaption[]{Temporal variation of density $n_{\rm H}$ in infalling
fluid elements that migrate from $8.2\times 10^3$ AU to $1.0\times10^3$ AU
({\em solid line}), $1.9\times10^4$ AU to $5.0\times10^3$ AU ({\em dotted
line}), $2.6\times10^4$ AU to $9.0\times10^3$ AU ({\em dashed line}),
and $3.1\times10^4$ AU to $1.3\times10^4$ AU ({\em dot-dashed line}) in
$2.00\times 10^5$ yr.
\label{fig:lagrange}}

\figcaption[]{Evolution of molecular abundances in a fluid element that
migrates from $8.2\times 10^3$ AU to $1.0\times 10^3$ AU in
$2.00\times 10^5$ yr, while the gas density ($n_{\rm H}$) varies from $1.7
\times 10^4$ cm$^{-3}$ to $5.8\times 10^6$ cm$^{-3}$.
\label{fig:evol}}

\figcaption[]{Distribution of molecular abundances at (a) $t=1.52\times 10^5$
yr, (b) $t=1.89\times 10^5$ yr, and (c) $t=2.00\times 10^5$ yr in a core with
the Larson-Penston flow. The central density ($n_{\rm H}$) of the core is
$3\times 10^5$ cm$^{-3}$ (a), $3\times 10^6$ cm$^{-3}$ (b), and $3\times 10^7$
cm$^{-3}$ (c).
\label{fig:dist}}

\figcaption[]{Molecular column densities vs distance from the center
in a Larson-Penston core at
$t=1.52\times 10^5$ yr ({\em dotted lines}), $1.89\times 10^5$ yr
({\em solid lines}), and $2.00\times 10^5$ yr ({\em dashed lines}).
\label{fig:column}}

\figcaption[]{Molecular evolution in a fluid element that migrates from
$1.5\times 10^4$ AU to $3.0\times 10^3$ AU in (a) $2.00\times 10^5$ yr,
and (b) $2.00 \times 10^6$ yr. The Larson-Penston collapse is assumed
for (a), while the collapse is slowed down by a factor $f$ of 10 in (b).
\label{fig:evol_slow}}

\figcaption[]{Molecular evolution in a fluid element as a function of
gas density. Thick solid lines show the time, with labels on the
right hand side of the figure, as a function of density.
Other details are the same as in Figure \ref{fig:evol_slow}.
\label{fig:evol_slow_R}}

\figcaption[]{Molecular distributions in cores with slow collapse.
The collapse time scale is larger by a factor of $f=3$  ({\em left panel})
and 10 ({\em right panel})
compared with the Larson-Penston solution. The age of the core is
(a) $t=f\times 1.52\times 10^5$ yr, (b) $f\times 1.89\times 10^5$ yr, and
(c) $f\times 2.00\times 10^5$ yr.
\label{fig:dist_slow}}

\figcaption[]{Column densities of assorted species at
$t=f\times 1.89\times 10^5$ yr plotted vs $R$. The solid lines are
for the Larson-Penston collapse ($f=1$), the dotted lines are for slow
collapse with $f=3$, and the dashed lines are for $f=10$.
The sticking probability is assumed
to be $S=1.0$ for these lines. The dot-dashed lines are for the case
of $f=3$, but with lower sticking probability $S=0.33$.
\label{fig:column_slow}}

\figcaption[]{Column densities of assorted species for models with
different initial conditions plotted vs $R$. The solid lines are
the same as those of Figure \ref{fig:column_slow}.
For the dotted lines, the initial
central density is $n_{\rm H}= 2.0\times 10^2$ cm$^{-3}$.
For the dashed lines, the gas has a pre-collapse phase of
$3\times 10^5$ yr
in a cloud of constant density $n_{\rm H}=2.0\times 10^4$ cm$^{-3}$.
The physical structure of the three core
models is the same, with the eventual central density
$n_{\rm H}= 3\times 10^6$
cm$^{-3}$.
\label{fig:column_init}}

\figcaption[]{Column densities at $t=1.89\times 10^5$ yr
plotted vs $R$ for models
with different adsorption energies: Model A ({\em solid lines}), Model B
({\em dotted lines}), and Model C ({\em dashed lines}). The Larson-Penston
collapse ($f=1$) is assumed. Descriptions of the models are given
in \S 3.6.
\label{fig:column_Eads}}

\figcaption[]{The column density ratio of DCO$^+$ to HCO$^+$ plotted vs $R$.
(a) The Larson-Penston core with $S=1.0$ at $t=1.52\times 10^5$
yr ({\em dotted line}), $1.89\times 10^5$ ({\em solid line}), and $2.00 \times
10^5$ ({\em dashed line}). (b) Thick lines for
$t=f\times 1.89\times 10^5$ yr; $f=3$ for the dotted
line, and $f=10$ for the dashed line, while $f=1$ for the solid line.
Thin lines for $t= 1.89\times 10^5$ yr with different
sets of adsorption energies;
the dotted line for Model B, and the dashed line for Model C
(see \S 3.6).
\label{fig:D}}

\clearpage
\begin{deluxetable}{l c l c}
\tablecaption{Initial Abundances
\tablenotemark{a}}
\tablewidth{0pt}
\tablehead{\colhead{Species} & \colhead{Abundance} &
\colhead{Species} & \colhead{Abundance} }
\startdata
   H$_2$  & 5.0(-1)\tablenotemark{b}   & HD     & 1.50(-5) \nl
   He     & 9.75(-2)  & N      & 2.47(-5) \nl
   O      & 1.80(-4)  & electrons & 7.87(-5) \nl
   C$^+$  & 7.86(-5)  & Na$^+$ & 2.25(-9) \nl
   Mg$^+$ & 1.09(-8)  & Si$^+$ & 9.74(-9) \nl
   P$^+$  & 2.16(-10) & S$^+$  & 9.14(-8) \nl
   Cl$^+$ & 1.00(-9)  & Fe$^+$ & 2.74(-9) \nl
   Grain  & 1.04(-2)\tablenotemark{c}  &        &          \nl
\enddata
\tablenotetext{a}{The abundances by number of the species relative to
hydrogen nuclei.}
\tablenotetext{b}{$a(b)$ means $a\times 10^b$.}
\tablenotetext{c}{The abundance of refractory grain cores by mass relative to
hydrogen.}
\label{tab:initial}
\end{deluxetable}

\begin{small}
\begin{deluxetable}{l c c c c c c c}
\tablecaption{Column Density [cm$^{-2}$] of Molecules in Larson-Penston Core at
$t=1.89\times 10^5$ yr as a Function of Distance from the Center}
\tablewidth{0pt}
\tablehead{\colhead{Species} & \colhead{1000 AU} & \colhead{3000 AU} &
\colhead{5000 AU}&
\colhead{7000 AU}& \colhead{9000 AU}& \colhead{11000 AU}& \colhead{13000 AU}}
\startdata
   Gaseous Species     &            &            &            &            &
&            &            \nl
    C        & 2.29E+17 & 2.50E+17 & 2.77E+17 & 2.74E+17 & 2.53E+17 &
2.19E+17 & 1.66E+17  \nl
    Cl       & 1.93E+13 & 2.07E+13 & 1.95E+13 & 1.61E+13 & 1.25E+13 &
9.16E+12 & 5.83E+12  \nl
    Fe       & 7.84E+13 & 7.86E+13 & 6.67E+13 & 4.97E+13 & 3.41E+13 &
2.05E+13 & 9.25E+12  \nl
    H        & 8.87E+19 & 9.11E+19 & 9.66E+19 & 1.08E+20 & 1.28E+20 &
1.58E+20 & 1.76E+20  \nl
    D        & 1.25E+16 & 1.08E+16 & 1.02E+16 & 9.96E+15 & 9.53E+15 &
8.73E+15 & 7.08E+15  \nl
    He       & 1.67E+22 & 8.32E+21 & 4.79E+21 & 3.08E+21 & 2.06E+21 &
1.36E+21 & 8.09E+20  \nl
    Mg       & 1.86E+14 & 2.05E+14 & 2.01E+14 & 1.69E+14 & 1.30E+14 &
9.05E+13 & 5.10E+13  \nl
    N        & 1.84E+18 & 1.37E+18 & 9.59E+17 & 6.68E+17 & 4.66E+17 &
3.17E+17 & 1.91E+17  \nl
    Na       & 3.78E+13 & 4.16E+13 & 4.02E+13 & 3.34E+13 & 2.57E+13 &
1.86E+13 & 1.12E+13  \nl
    O        & 8.39E+18 & 6.29E+18 & 4.48E+18 & 3.17E+18 & 2.26E+18 &
1.57E+18 & 9.77E+17  \nl
    S        & 1.85E+15 & 1.93E+15 & 1.77E+15 & 1.43E+15 & 1.10E+15 &
7.88E+14 & 4.88E+14  \nl
    Si       & 1.53E+14 & 1.69E+14 & 1.66E+14 & 1.38E+14 & 1.05E+14 &
7.25E+13 & 4.24E+13  \nl
    C$_2$       & 8.85E+13 & 9.28E+13 & 9.93E+13 & 1.07E+14 & 1.18E+14 &
1.32E+14 & 1.39E+14  \nl
    CH       & 1.50E+13 & 1.58E+13 & 1.76E+13 & 1.92E+13 & 2.00E+13 &
1.96E+13 & 1.63E+13  \nl
    CN       & 9.00E+14 & 9.96E+14 & 1.06E+15 & 9.83E+14 & 8.55E+14 &
6.98E+14 & 4.82E+14  \nl
    CO       & 1.59E+18 & 1.70E+18 & 1.55E+18 & 1.21E+18 & 8.92E+17 &
6.11E+17 & 3.58E+17  \nl
    CS       & 5.03E+13 & 5.48E+13 & 5.93E+13 & 5.84E+13 & 5.44E+13 &
4.90E+13 & 4.26E+13  \nl
    H$_2$       & 8.54E+22 & 4.26E+22 & 2.45E+22 & 1.57E+22 & 1.05E+22 &
6.91E+21 & 4.05E+21  \nl
    HD       & 2.50E+18 & 1.25E+18 & 7.15E+17 & 4.57E+17 & 3.04E+17 &
2.00E+17 & 1.17E+17  \nl
    N$_2$       & 1.63E+16 & 1.08E+16 & 6.66E+15 & 4.11E+15 & 2.53E+15 &
1.45E+15 & 6.74E+14  \nl
    NO       & 9.93E+14 & 5.36E+14 & 3.38E+14 & 2.36E+14 & 1.71E+14 &
1.21E+14 & 7.24E+13  \nl
    O$_2$       & 3.00E+16 & 3.37E+15 & 8.46E+14 & 4.41E+14 & 2.80E+14 &
1.85E+14 & 1.10E+14  \nl
    OH       & 6.10E+13 & 3.96E+13 & 3.06E+13 & 2.56E+13 & 2.19E+13 &
1.83E+13 & 1.35E+13  \nl
    SiO      & 1.94E+13 & 1.78E+13 & 1.59E+13 & 1.47E+13 & 1.43E+13 &
1.40E+13 & 1.05E+13  \nl
    C$_2$H      & 1.93E+14 & 2.05E+14 & 2.31E+14 & 2.53E+14 & 2.58E+14 &
2.40E+14 & 1.77E+14  \nl
    C$_2$N      & 6.21E+13 & 6.73E+13 & 7.53E+13 & 7.93E+13 & 7.49E+13 &
5.85E+13 & 2.95E+13  \nl
    C$_2$S      & 2.54E+13 & 2.74E+13 & 3.17E+13 & 3.39E+13 & 3.11E+13 &
2.36E+13 & 1.24E+13  \nl
    C$_3$       & 5.24E+13 & 5.43E+13 & 5.84E+13 & 6.52E+13 & 7.46E+13 &
8.57E+13 & 9.15E+13  \nl
    CH$_2$      & 2.29E+15 & 2.53E+15 & 2.68E+15 & 2.46E+15 & 2.10E+15 &
1.67E+15 & 1.11E+15  \nl
    CHD      & 9.63E+12 & 1.02E+13 & 1.02E+13 & 9.46E+12 & 8.32E+12 &
6.80E+12 & 4.25E+12  \nl
    CO$_2$      & 3.76E+15 & 3.47E+15 & 1.82E+15 & 5.77E+14 & 1.61E+14 &
4.64E+13 & 1.07E+13  \nl
    H$_2$O      & 2.72E+16 & 2.26E+16 & 1.75E+16 & 1.18E+16 & 6.84E+15 &
3.10E+15 & 8.62E+14  \nl
    HDO      & 9.57E+14 & 5.59E+14 & 3.57E+14 & 2.15E+14 & 1.12E+14 &
4.44E+13 & 1.01E+13  \nl
    HCN      & 2.05E+15 & 2.34E+15 & 2.43E+15 & 1.93E+15 & 1.34E+15 &
7.76E+14 & 2.64E+14  \nl
    DCN      & 1.58E+13 & 1.75E+13 & 1.68E+13 & 1.30E+13 & 9.05E+12 &
5.12E+12 & 1.59E+12  \nl
    HNC      & 9.15E+14 & 1.05E+15 & 1.08E+15 & 8.28E+14 & 5.32E+14 &
2.58E+14 & 5.71E+13  \nl
    DNC      & 2.00E+13 & 2.26E+13 & 2.09E+13 & 1.46E+13 & 8.92E+12 &
4.01E+12 & 6.45E+11  \nl
    NH$_2$      & 6.49E+12 & 6.00E+12 & 5.57E+12 & 4.91E+12 & 4.04E+12 &
2.90E+12 & 1.49E+12  \nl
    OCN      & 6.43E+14 & 4.15E+14 & 2.40E+14 & 1.26E+14 & 5.65E+13 &
2.35E+13 & 7.91E+12  \nl
    OCS      & 3.77E+13 & 4.44E+13 & 4.28E+13 & 2.76E+13 & 1.31E+13 &
4.69E+12 & 1.20E+12  \nl
    C$_2$H$_2$     & 2.94E+13 & 2.49E+13 & 2.15E+13 & 1.97E+13 &
1.79E+13 & 1.55E+13 & 1.17E+13  \nl
    C$_3$H      & 2.44E+14 & 2.61E+14 & 2.93E+14 & 3.20E+14 & 3.23E+14 &
2.81E+14 & 1.66E+14  \nl
    C$_3$N      & 4.71E+13 & 5.06E+13 & 5.72E+13 & 6.19E+13 & 6.06E+13 &
5.03E+13 & 2.82E+13  \nl
    C$_4$       & 3.30E+13 & 3.40E+13 & 3.65E+13 & 4.08E+13 & 4.70E+13 &
5.46E+13 & 6.08E+13  \nl
    CH$_3$      & 7.95E+12 & 8.40E+12 & 8.99E+12 & 9.43E+12 & 9.53E+12 &
9.06E+12 & 7.27E+12  \nl
    H$_2$CO     & 4.32E+14 & 4.75E+14 & 5.13E+14 & 4.84E+14 & 3.85E+14 &
2.41E+14 & 9.73E+13  \nl
    HDCO     & 1.76E+13 & 1.91E+13 & 1.98E+13 & 1.81E+13 & 1.43E+13 &
8.92E+12 & 3.56E+12  \nl
    H$_2$CS     & 1.07E+13 & 1.24E+13 & 1.10E+13 & 7.12E+12 & 3.81E+12 &
1.64E+12 & 5.50E+11  \nl
    NH$_3$      & 3.31E+13 & 2.84E+13 & 2.27E+13 & 1.68E+13 & 1.15E+13 &
6.51E+12 & 2.44E+12  \nl
    C$_2$H$_2$N    & 7.30E+14 & 8.57E+14 & 8.68E+14 & 6.24E+14 &
3.48E+14 & 1.35E+14 & 2.26E+13  \nl
    C$_2$HDN    & 2.22E+13 & 2.52E+13 & 2.21E+13 & 1.43E+13 & 7.46E+12 &
2.77E+12 & 4.59E+11  \nl
    C$_2$H$_2$O    & 2.60E+14 & 2.95E+14 & 3.31E+14 & 2.95E+14 &
1.98E+14 & 9.40E+13 & 2.59E+13  \nl
    C$_2$H$_3$     & 1.44E+14 & 1.57E+14 & 1.78E+14 & 1.89E+14 &
1.72E+14 & 1.22E+14 & 5.98E+13  \nl
    C$_2$H$_2$D    & 8.38E+12 & 9.15E+12 & 1.03E+13 & 1.07E+13 &
9.58E+12 & 6.59E+12 & 2.97E+12  \nl
    C$_3$H$_2$     & 1.77E+13 & 1.87E+13 & 2.07E+13 & 2.33E+13 &
2.48E+13 & 2.36E+13 & 1.79E+13  \nl
    C$_4$H      & 1.36E+14 & 1.44E+14 & 1.63E+14 & 1.87E+14 & 1.88E+14 &
1.56E+14 & 1.04E+14  \nl
    C$_4$N      & 4.93E+12 & 5.29E+12 & 6.03E+12 & 6.58E+12 & 6.27E+12 &
4.91E+12 & 2.72E+12  \nl
    C$_5$       & 1.28E+13 & 1.32E+13 & 1.41E+13 & 1.59E+13 & 1.84E+13 &
2.12E+13 & 2.28E+13  \nl
    CH$_2$O$_2$    & 3.55E+13 & 3.60E+13 & 2.76E+13 & 1.58E+13 &
7.13E+12 & 2.19E+12 & 3.00E+11  \nl
    CH$_4$      & 4.01E+16 & 4.60E+16 & 4.54E+16 & 3.39E+16 & 2.08E+16 &
9.99E+15 & 3.00E+15  \nl
    CH$_3$D     & 3.69E+15 & 4.22E+15 & 4.08E+15 & 3.02E+15 & 1.85E+15 &
8.82E+14 & 2.60E+14  \nl
    HC$_3$N     & 1.62E+13 & 1.77E+13 & 2.05E+13 & 2.13E+13 & 1.86E+13 &
1.23E+13 & 4.43E+12  \nl
    NH$_2$CN    & 9.35E+12 & 1.08E+13 & 1.16E+13 & 9.39E+12 & 6.23E+12 &
3.05E+12 & 7.98E+11  \nl
    C$_2$H$_3$N    & 3.58E+13 & 3.94E+13 & 4.75E+13 & 4.82E+13 &
3.74E+13 & 1.89E+13 & 3.38E+12  \nl
    C$_4$H$_2$     & 2.55E+13 & 2.90E+13 & 3.15E+13 & 2.61E+13 &
1.46E+13 & 5.09E+12 & 1.00E+12  \nl
    C$_5$H      & 5.85E+12 & 6.15E+12 & 6.84E+12 & 7.86E+12 & 8.63E+12 &
7.95E+12 & 4.71E+12  \nl
    C$_5$N      & 4.92E+12 & 5.28E+12 & 5.99E+12 & 6.54E+12 & 6.29E+12 &
5.06E+12 & 3.01E+12  \nl
    CH$_4$O     & 1.73E+13 & 1.98E+13 & 2.29E+13 & 1.98E+13 & 1.19E+13 &
4.43E+12 & 6.63E+11  \nl
    C$_3$H$_4$     & 2.02E+13 & 2.29E+13 & 2.76E+13 & 2.48E+13 &
1.44E+13 & 4.80E+12 & 7.04E+11  \nl
    C$_5$H$_2$     & 7.00E+12 & 7.58E+12 & 8.60E+12 & 9.32E+12 &
8.73E+12 & 6.16E+12 & 2.18E+12  \nl
    HC$_5$N     & 1.64E+13 & 1.82E+13 & 2.13E+13 & 2.08E+13 & 1.51E+13 &
7.63E+12 & 2.02E+12  \nl
    C$_6$H$_2$     & 1.27E+13 & 1.38E+13 & 1.45E+13 & 1.35E+13 &
1.06E+13 & 6.81E+12 & 2.77E+12  \nl
    CH$_3$C$_4$H   & 3.97E+13 & 4.55E+13 & 4.96E+13 & 4.11E+13 &
2.55E+13 & 1.09E+13 & 2.17E+12  \nl
    e        & 4.19E+14 & 3.35E+14 & 2.78E+14 & 2.38E+14 & 2.06E+14 &
1.82E+14 & 1.64E+14  \nl
    C$^+$      & 4.53E+13 & 4.30E+13 & 4.05E+13 & 4.04E+13 & 4.26E+13 &
4.84E+13 & 6.25E+13  \nl
    Fe$^+$     & 1.16E+13 & 1.22E+13 & 1.26E+13 & 1.27E+13 & 1.29E+13 &
1.29E+13 & 1.15E+13  \nl
    Mg$^+$     & 1.96E+13 & 2.09E+13 & 2.26E+13 & 2.34E+13 & 2.39E+13 &
2.43E+13 & 2.31E+13  \nl
    Na$^+$     & 5.53E+12 & 6.06E+12 & 6.42E+12 & 6.09E+12 & 5.38E+12 &
4.54E+12 & 3.76E+12  \nl
    C$_2$N$^+$    & 4.53E+12 & 5.10E+12 & 5.27E+12 & 4.61E+12 & 3.91E+12
& 3.15E+12 & 1.85E+12  \nl
    H$_3$$^+$     & 1.65E+13 & 1.18E+13 & 9.29E+12 & 7.81E+12 & 6.60E+12
& 5.34E+12 & 3.73E+12  \nl
    HCO$^+$    & 6.88E+13 & 6.52E+13 & 5.41E+13 & 4.30E+13 & 3.33E+13 &
2.35E+13 & 1.23E+13  \nl
    H$_2$CN$^+$   & 1.68E+13 & 1.93E+13 & 1.90E+13 & 1.39E+13 & 8.74E+12
& 4.30E+12 & 1.05E+12  \nl
    H$_3$O$^+$    & 1.59E+14 & 9.50E+13 & 6.16E+13 & 4.25E+13 & 2.90E+13
& 1.80E+13 & 8.09E+12  \nl
   Mantle Species   &            &            &            &            &
&            &            \nl
    C        & 7.60E+16 & 7.05E+16 & 5.36E+16 & 3.62E+16 & 2.43E+16 &
1.63E+16 & 9.93E+15  \nl
    Cl       & 1.44E+14 & 5.97E+13 & 2.77E+13 & 1.45E+13 & 8.14E+12 &
4.63E+12 & 2.41E+12  \nl
    Fe       & 3.71E+14 & 1.35E+14 & 5.31E+13 & 2.29E+13 & 1.00E+13 &
4.07E+12 & 1.32E+12  \nl
    Mg       & 1.64E+15 & 6.79E+14 & 3.04E+14 & 1.48E+14 & 7.42E+13 &
3.58E+13 & 1.46E+13  \nl
    N        & 2.10E+18 & 6.07E+17 & 2.07E+17 & 8.75E+16 & 4.23E+16 &
2.16E+16 & 1.04E+16  \nl
    Na       & 3.36E+14 & 1.39E+14 & 6.24E+13 & 3.10E+13 & 1.61E+13 &
8.19E+12 & 3.48E+12  \nl
    O        & 9.26E+18 & 2.70E+18 & 9.41E+17 & 4.08E+17 & 2.02E+17 &
1.06E+17 & 5.29E+16  \nl
    P        & 2.85E+13 & 1.21E+13 & 5.71E+12 & 3.02E+12 & 1.70E+12 &
9.53E+11 & 4.73E+11  \nl
    S        & 1.17E+16 & 4.81E+15 & 2.19E+15 & 1.12E+15 & 6.15E+14 &
3.36E+14 & 1.62E+14  \nl
    Si       & 1.36E+15 & 5.62E+14 & 2.52E+14 & 1.23E+14 & 6.33E+13 &
3.23E+13 & 1.46E+13  \nl
    C$_2$       & 4.27E+15 & 2.29E+15 & 1.43E+15 & 9.86E+14 & 7.08E+14 &
4.99E+14 & 3.12E+14  \nl
    CN       & 1.25E+16 & 5.64E+15 & 2.87E+15 & 1.63E+15 & 9.71E+14 &
5.67E+14 & 2.91E+14  \nl
    CO       & 1.06E+19 & 4.12E+18 & 1.75E+18 & 8.33E+17 & 4.25E+17 &
2.18E+17 & 1.02E+17  \nl
    CS       & 9.53E+14 & 4.06E+14 & 2.03E+14 & 1.18E+14 & 7.63E+13 &
5.25E+13 & 3.46E+13  \nl
    N$_2$       & 5.70E+16 & 1.46E+16 & 4.67E+15 & 1.87E+15 & 8.22E+14 &
3.49E+14 & 1.23E+14  \nl
    NO       & 3.27E+15 & 8.99E+14 & 3.50E+14 & 1.76E+14 & 9.63E+13 &
5.09E+13 & 2.21E+13  \nl
    O$_2$       & 4.39E+16 & 2.92E+15 & 6.12E+14 & 2.71E+14 & 1.44E+14 &
7.78E+13 & 3.63E+13  \nl
    OH       & 3.95E+14 & 1.56E+14 & 7.99E+13 & 4.83E+13 & 3.06E+13 &
1.87E+13 & 9.71E+12  \nl
    SiO      & 9.70E+13 & 4.53E+13 & 2.57E+13 & 1.64E+13 & 1.05E+13 &
5.95E+12 & 2.42E+12  \nl
    SO       & 3.07E+13 & 8.13E+12 & 1.88E+12 & 3.88E+11 & 8.35E+10 &
1.91E+10 & 3.66E+09  \nl
    C$_2$H      & 5.47E+15 & 2.46E+15 & 1.26E+15 & 7.05E+14 & 4.02E+14 &
2.16E+14 & 9.44E+13  \nl
    C$_2$N      & 1.14E+15 & 4.99E+14 & 2.35E+14 & 1.16E+14 & 5.45E+13 &
2.23E+13 & 6.46E+12  \nl
    C$_2$S      & 3.36E+14 & 1.40E+14 & 6.54E+13 & 3.28E+13 & 1.61E+13 &
7.10E+12 & 2.46E+12  \nl
    C$_3$       & 1.93E+15 & 1.06E+15 & 6.63E+14 & 4.58E+14 & 3.28E+14 &
2.29E+14 & 1.41E+14  \nl
    CH$_2$      & 1.44E+16 & 6.99E+15 & 3.60E+15 & 2.00E+15 & 1.17E+15 &
6.73E+14 & 3.33E+14  \nl
    CHD      & 4.52E+13 & 2.28E+13 & 1.25E+13 & 7.47E+12 & 4.53E+12 &
2.53E+12 & 1.05E+12  \nl
    CO$_2$      & 3.22E+16 & 6.92E+15 & 1.22E+15 & 2.10E+14 & 4.17E+13 &
8.88E+12 & 1.50E+12  \nl
    H$_2$O      & 2.55E+17 & 8.22E+16 & 2.82E+16 & 1.02E+16 & 3.51E+15 &
1.02E+15 & 2.09E+14  \nl
    HDO      & 5.62E+15 & 1.52E+15 & 4.63E+14 & 1.52E+14 & 4.68E+13 &
1.17E+13 & 1.90E+12  \nl
    HCN      & 2.71E+16 & 9.82E+15 & 3.78E+15 & 1.55E+15 & 6.21E+14 &
2.11E+14 & 4.58E+13  \nl
    DCN      & 1.60E+14 & 5.92E+13 & 2.34E+13 & 9.89E+12 & 4.04E+12 &
1.34E+12 & 2.75E+11  \nl
    HCO      & 1.09E+14 & 3.19E+13 & 8.29E+12 & 1.87E+12 & 4.22E+11 &
1.13E+11 & 3.22E+10  \nl
    HNC      & 1.21E+16 & 4.31E+15 & 1.58E+15 & 5.99E+14 & 2.11E+14 &
5.67E+13 & 8.25E+12  \nl
    DNC      & 2.01E+14 & 6.85E+13 & 2.39E+13 & 8.66E+12 & 2.93E+12 &
7.03E+11 & 6.77E+10  \nl
    HNO      & 5.49E+13 & 1.92E+13 & 7.45E+12 & 3.10E+12 & 1.26E+12 &
4.57E+11 & 1.27E+11  \nl
    N$_2$O      & 7.35E+13 & 1.95E+13 & 5.38E+12 & 1.61E+12 & 4.85E+11 &
1.33E+11 & 2.91E+10  \nl
    OCN      & 2.39E+15 & 6.34E+14 & 1.88E+14 & 5.92E+13 & 1.86E+13 &
5.68E+12 & 1.35E+12  \nl
    OCS      & 3.17E+14 & 1.03E+14 & 3.23E+13 & 9.72E+12 & 2.70E+12 &
6.69E+11 & 1.30E+11  \nl
    C$_2$H$_2$     & 2.08E+14 & 9.40E+13 & 5.17E+13 & 3.25E+13 &
2.11E+13 & 1.33E+13 & 7.31E+12  \nl
    C$_3$H      & 5.16E+15 & 2.33E+15 & 1.17E+15 & 6.15E+14 & 3.13E+14 &
1.39E+14 & 4.48E+13  \nl
    C$_3$D      & 5.88E+13 & 2.66E+13 & 1.31E+13 & 6.76E+12 & 3.27E+12 &
1.31E+12 & 3.28E+11  \nl
    C$_3$N      & 7.71E+14 & 3.40E+14 & 1.65E+14 & 8.42E+13 & 4.17E+13 &
1.82E+13 & 5.77E+12  \nl
    C$_3$S      & 4.03E+13 & 1.59E+13 & 7.22E+12 & 3.66E+12 & 1.90E+12 &
9.62E+11 & 4.24E+11  \nl
    C$_4$       & 9.71E+14 & 5.32E+14 & 3.35E+14 & 2.34E+14 & 1.71E+14 &
1.24E+14 & 7.97E+13  \nl
    CH$_3$      & 1.33E+14 & 7.14E+13 & 4.37E+13 & 2.92E+13 & 1.99E+13 &
1.31E+13 & 7.42E+12  \nl
    H$_2$CO     & 6.20E+15 & 2.56E+15 & 1.13E+15 & 5.17E+14 & 2.25E+14 &
8.41E+13 & 2.35E+13  \nl
    HDCO     & 2.04E+14 & 8.47E+13 & 3.80E+13 & 1.77E+13 & 7.85E+12 &
2.97E+12 & 8.14E+11  \nl
    H$_2$CS     & 5.15E+13 & 1.88E+13 & 6.77E+12 & 2.49E+12 & 8.81E+11 &
2.83E+11 & 7.36E+10  \nl
    NH$_3$      & 3.09E+14 & 1.13E+14 & 4.62E+13 & 2.07E+13 & 9.03E+12 &
3.39E+12 & 9.33E+11  \nl
    C$_2$H$_2$N    & 8.63E+15 & 2.95E+15 & 1.01E+15 & 3.41E+14 &
1.00E+14 & 2.11E+13 & 2.07E+12  \nl
    C$_2$HDN    & 1.91E+14 & 6.21E+13 & 2.02E+13 & 6.55E+12 & 1.89E+12 &
3.98E+11 & 3.87E+10  \nl
    C$_2$H$_2$O    & 3.16E+15 & 1.22E+15 & 4.87E+14 & 1.92E+14 &
6.71E+13 & 1.87E+13 & 3.25E+12  \nl
    C$_2$HDO    & 1.00E+14 & 3.87E+13 & 1.54E+13 & 6.06E+12 & 2.10E+12 &
5.66E+11 & 8.75E+10  \nl
    C$_2$H$_3$     & 2.37E+15 & 1.07E+15 & 5.29E+14 & 2.70E+14 &
1.30E+14 & 5.45E+13 & 1.71E+13  \nl
    C$_2$H$_2$D    & 1.20E+14 & 5.41E+13 & 2.65E+13 & 1.34E+13 &
6.33E+12 & 2.49E+12 & 6.79E+11  \nl
    C$_3$H$_2$     & 3.88E+14 & 1.89E+14 & 1.02E+14 & 5.90E+13 &
3.45E+13 & 1.92E+13 & 9.05E+12  \nl
    C$_4$H      & 4.22E+15 & 1.74E+15 & 7.88E+14 & 3.91E+14 & 2.00E+14 &
9.98E+13 & 4.35E+13  \nl
    C$_4$N      & 1.05E+14 & 4.56E+13 & 2.11E+13 & 1.01E+13 & 4.58E+12 &
1.86E+12 & 5.84E+11  \nl
    C$_5$       & 5.44E+14 & 2.83E+14 & 1.68E+14 & 1.11E+14 & 7.71E+13 &
5.37E+13 & 3.45E+13  \nl
    CH$_2$O$_2$    & 2.04E+14 & 6.44E+13 & 1.97E+13 & 5.87E+12 &
1.52E+12 & 2.81E+11 & 2.46E+10  \nl
    CH$_4$      & 4.24E+17 & 1.72E+17 & 6.99E+16 & 2.82E+16 & 1.05E+16 &
3.23E+15 & 6.55E+14  \nl
    CH$_3$D     & 3.65E+16 & 1.47E+16 & 5.93E+15 & 2.39E+15 & 8.92E+14 &
2.71E+14 & 5.32E+13  \nl
    HC$_3$N     & 2.55E+14 & 1.05E+14 & 4.64E+13 & 2.10E+13 & 8.77E+12 &
2.94E+12 & 6.06E+11  \nl
    NH$_2$CN    & 1.49E+14 & 5.59E+13 & 2.19E+13 & 8.67E+12 & 3.10E+12 &
8.46E+11 & 1.38E+11  \nl
    C$_2$H$_3$N    & 8.04E+14 & 2.88E+14 & 1.11E+14 & 4.30E+13 &
1.46E+13 & 3.39E+12 & 3.23E+11  \nl
    C$_3$H$_3$     & 4.80E+13 & 2.18E+13 & 1.01E+13 & 4.38E+12 &
1.60E+12 & 4.01E+11 & 4.59E+10  \nl
    C$_4$H$_2$     & 2.46E+14 & 1.03E+14 & 4.14E+13 & 1.47E+13 &
4.20E+12 & 9.00E+11 & 1.33E+11  \nl
    C$_5$H      & 2.51E+14 & 1.07E+14 & 4.90E+13 & 2.35E+13 & 1.08E+13 &
4.38E+12 & 1.38E+12  \nl
    C$_5$N      & 9.66E+13 & 4.25E+13 & 2.00E+13 & 9.78E+12 & 4.63E+12 &
1.99E+12 & 6.81E+11  \nl
    C$_6$       & 2.29E+14 & 1.13E+14 & 6.31E+13 & 3.83E+13 & 2.39E+13 &
1.48E+13 & 8.49E+12  \nl
    CH$_4$O     & 3.57E+14 & 1.30E+14 & 4.68E+13 & 1.58E+13 & 4.45E+12 &
8.66E+11 & 7.99E+10  \nl
    C$_3$H$_4$     & 2.70E+14 & 1.12E+14 & 4.48E+13 & 1.57E+13 &
4.29E+12 & 7.82E+11 & 7.21E+10  \nl
    C$_3$H$_3$D    & 1.96E+13 & 8.14E+12 & 3.24E+12 & 1.13E+12 &
3.05E+11 & 5.38E+10 & 4.53E+09  \nl
    C$_5$H$_2$     & 1.28E+14 & 5.84E+13 & 2.78E+13 & 1.30E+13 &
5.31E+12 & 1.64E+12 & 2.68E+11  \nl
    C$_6$H      & 1.89E+14 & 7.47E+13 & 3.16E+13 & 1.37E+13 & 5.55E+12 &
1.85E+12 & 4.22E+11  \nl
    C$_7$       & 1.00E+14 & 4.83E+13 & 2.59E+13 & 1.50E+13 & 8.78E+12 &
5.00E+12 & 2.66E+12  \nl
    HC$_5$N     & 2.79E+14 & 1.06E+14 & 4.10E+13 & 1.54E+13 & 5.10E+12 &
1.31E+12 & 2.00E+11  \nl
    C$_6$H$_2$     & 1.35E+14 & 6.01E+13 & 2.76E+13 & 1.26E+13 &
5.25E+12 & 1.77E+12 & 3.63E+11  \nl
    C$_7$H      & 1.21E+14 & 4.43E+13 & 1.71E+13 & 6.76E+12 & 2.46E+12 &
7.05E+11 & 1.05E+11  \nl
    C$_8$       & 4.59E+13 & 2.10E+13 & 1.04E+13 & 5.43E+12 & 2.77E+12 &
1.30E+12 & 5.38E+11  \nl
    C$_7$H$_2$     & 5.20E+13 & 2.05E+13 & 8.21E+12 & 3.15E+12 &
1.05E+12 & 2.65E+11 & 4.12E+10  \nl
    C$_8$H      & 1.08E+14 & 4.02E+13 & 1.58E+13 & 6.36E+12 & 2.38E+12 &
7.34E+11 & 1.42E+11  \nl
    CH$_3$C$_4$H & 3.60E+14 & 1.49E+14 & 6.09E+13 & 2.33E+13 &
7.44E+12 & 1.68E+12 & 1.79E+11  \nl
    HC$_7$N     & 5.05E+13 & 1.86E+13 & 7.06E+12 & 2.62E+12 & 8.48E+11 &
1.97E+11 & 1.97E+10  \nl
    C$_8$H$_2$     & 6.27E+13 & 2.54E+13 & 1.04E+13 & 4.05E+12 &
1.35E+12 & 3.23E+11 & 3.76E+10  \nl
    C$_9$H$_2$     & 3.04E+13 & 1.04E+13 & 3.51E+12 & 1.12E+12 &
2.92E+11 & 5.20E+10 & 3.68E+09  \nl
    HC$_9$N     & 2.16E+13 & 7.39E+12 & 2.55E+12 & 8.33E+11 & 2.27E+11 &
4.19E+10 & 3.07E+09  \nl
\enddata
\label{tab:elect}
\end{deluxetable}
\end{small}

\begin{deluxetable}{l c c}
\tablecaption{Adsorption Energies (K)}
\tablewidth{0pt}
\tablehead{\colhead{Species} & \colhead{Model A\tablenotemark{a}} &
\colhead{Model B\tablenotemark{b}} }
\startdata
    CO         &    1210    &      960   \nl
    N$_2$      &    1210    &      750   \nl
    CO$_2$     &    2500    &     2690   \nl
    H$_2$O     &    1860    &     4820   \nl
    HCN        &    1760    &     4280   \nl
    SO$_2$     &    3070    &     3460   \nl
    C$_2$H$_2$ &    1610    &     2490   \nl
    NH$_3$     &    1110    &     3080   \nl
    CH$_4$     &    1360    &     1120   \nl
\enddata
\tablenotetext{a}{Hasegawa \& Herbst 1993}
\tablenotetext{b}{See text for references.}
\label{tab:adsorb}
\end{deluxetable}

\begin{deluxetable}{c c c c}
\tablecaption{Molecular Column Densities in Collapsing Core}
\tablewidth{0pt}
\tablehead{\colhead{} & \colhead{CCS} &
\colhead{CO} & \colhead{N$_2$H$^+$} }
\startdata
    L1544   & $N_{\rm peak}=4\times 10^{13}$ cm$^{-2}$~ \tablenotemark{a} &
              $N_{\rm peak}=2\times 10^{18}$ cm$^{-2}$~ \tablenotemark{b} &
              $N_{\rm peak}=6\times 10^{12}$ cm$^{-2}$~ \tablenotemark{c}  \nl
            & $R_{\rm hole}\sim 7500$ AU \tablenotemark{a}          &
              $R_{\rm hole}\sim 6500$ AU \tablenotemark{d}          &
              centrally peaked                                       \nl
            &     &       &                                          \nl
f=1,S=1.0& $N_{\rm peak}=3.3\times 10^{13}$ cm$^{-2}$            &
              $N_{\rm peak}=1.7\times 10^{18}$ cm$^{-2}$            &
              $N_{\rm peak}=2.5\times 10^{11}$ cm$^{-2}$             \nl
            & $R_{\rm hole}\sim 7000$ AU                            &
              $R_{\rm hole}\sim 4000$ AU                            &
              centrally peaked                                       \nl
            &                    &             &                     \nl
f=3,S=1.0& $N_{\rm peak}=1.4\times 10^{12}$ cm$^{-2}$            &
              $N_{\rm peak}=4.7\times 10^{17}$ cm$^{-2}$            &
              $N_{\rm peak}=1.4\times 10^{12}$ cm$^{-2}$             \nl
            & $R_{\rm hole}=13000$ AU                               &
              $R_{\rm hole}=7500$ AU                                &
              centrally peaked                                       \nl
            &                    &             &                     \nl
f=10,S=1.0&$N_{\rm peak}=8.0\times 10^9$ cm$^{-2}$               &
              $N_{\rm peak}=9.5\times 10^{16}$ cm$^{-2}$             &
              $N_{\rm peak}=2.4\times 10^{12}$ cm$^{-2}$              \nl
            & $R_{\rm hole}=14000$ AU                               &
              centrally peaked                                      &
              centrally peaked                                       \nl
            &                    &             &                     \nl
f=3,S=0.3& $N_{\rm peak}=4.3\times 10^{12}$ cm$^{-2}$             &
              $N_{\rm peak}=1.7\times 10^{18}$ cm$^{-2}$             &
              $N_{\rm peak}=9.9\times 10^{11}$ cm$^{-2}$              \nl
            & $R_{\rm hole}=12000$ AU                               &
              $R_{\rm hole}=4000$ AU                                &
              centrally peaked                                       \nl
\enddata
\tablenotetext{a}{Ohashi et al. 1999}
\tablenotetext{b}{Caselli 2000, private communication}
\tablenotetext{c}{Benson et al. 1998}
\tablenotetext{d}{Caselli et al. 1999}
\label{tab:comp}
\end{deluxetable}

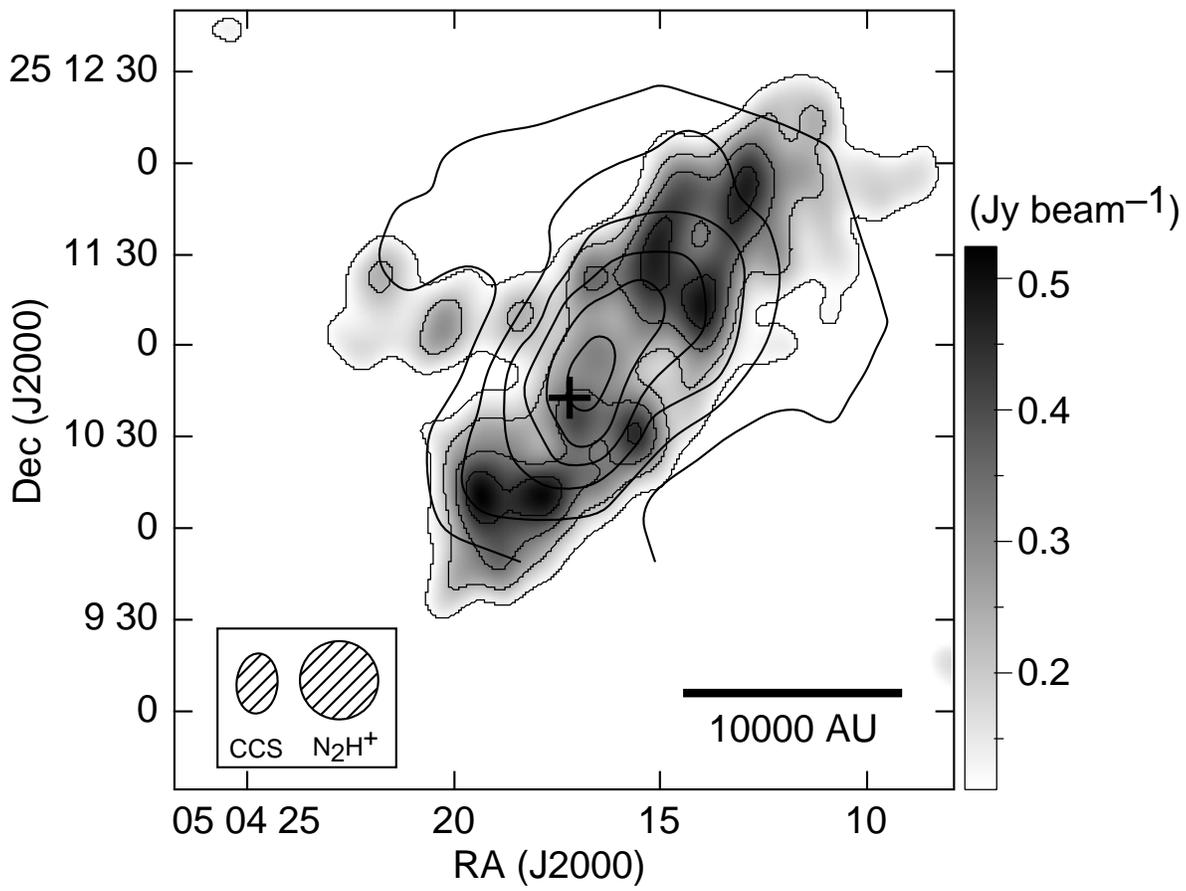
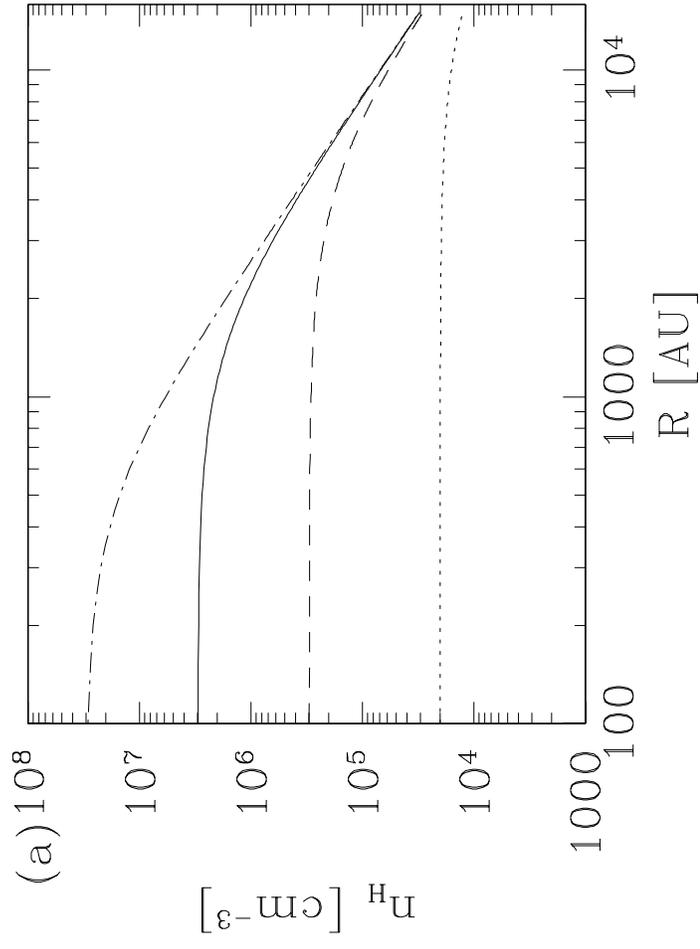
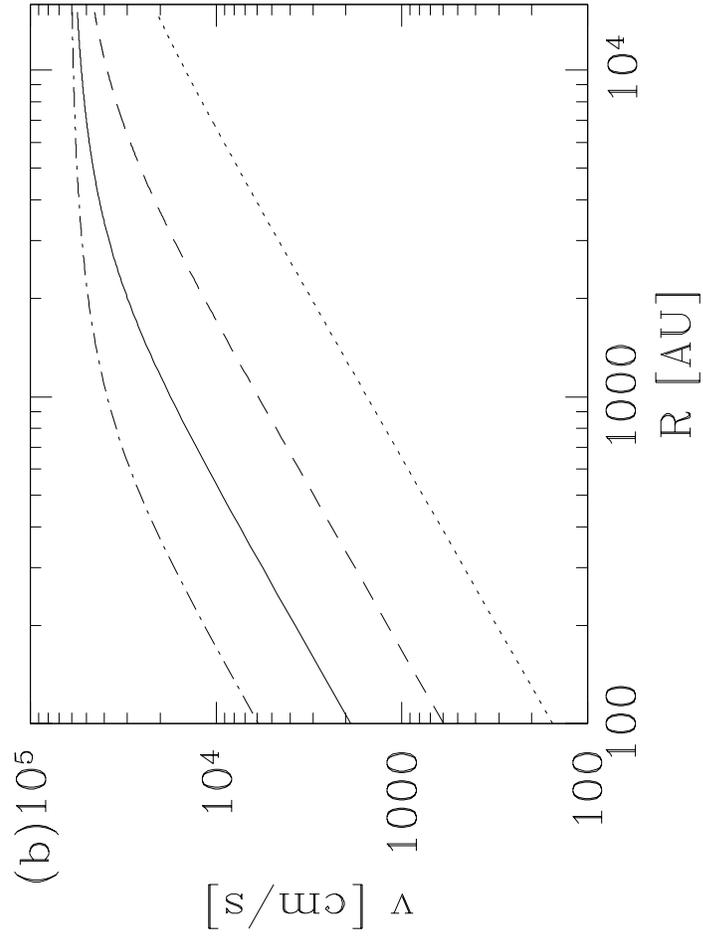
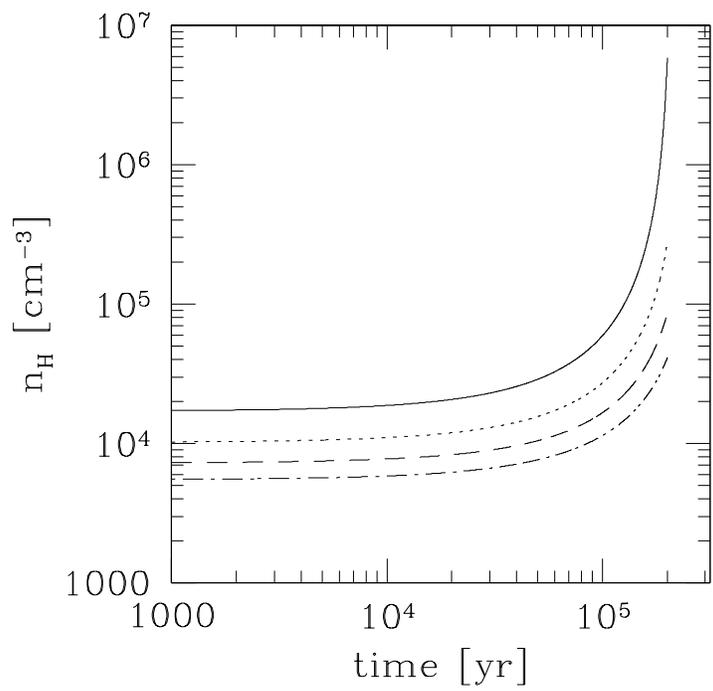
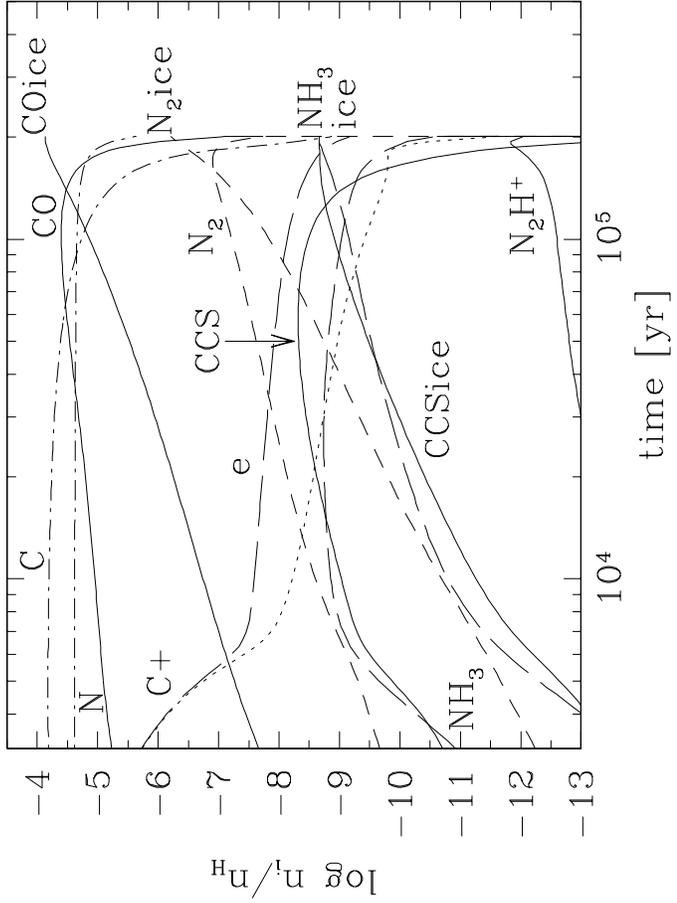
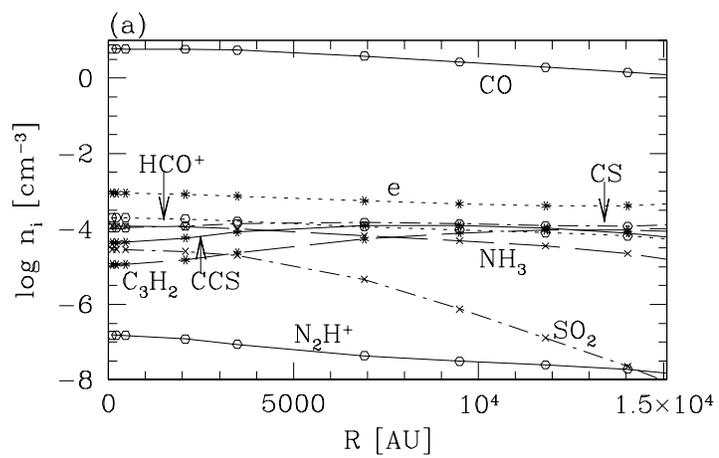
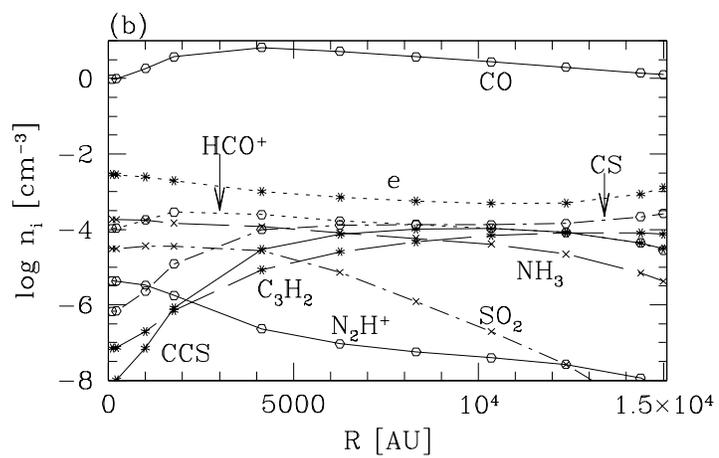
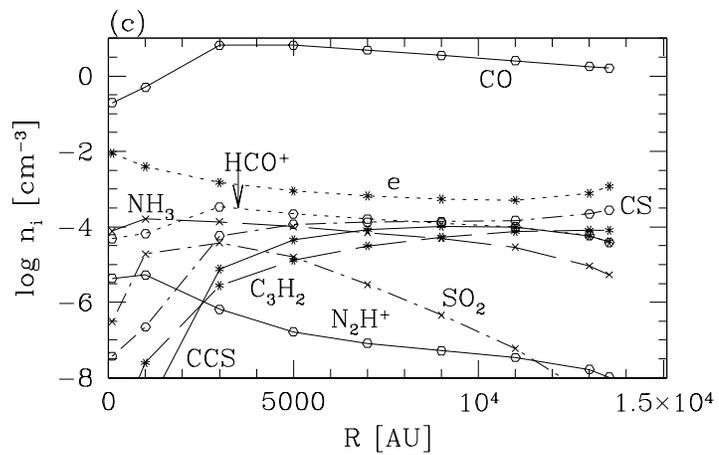
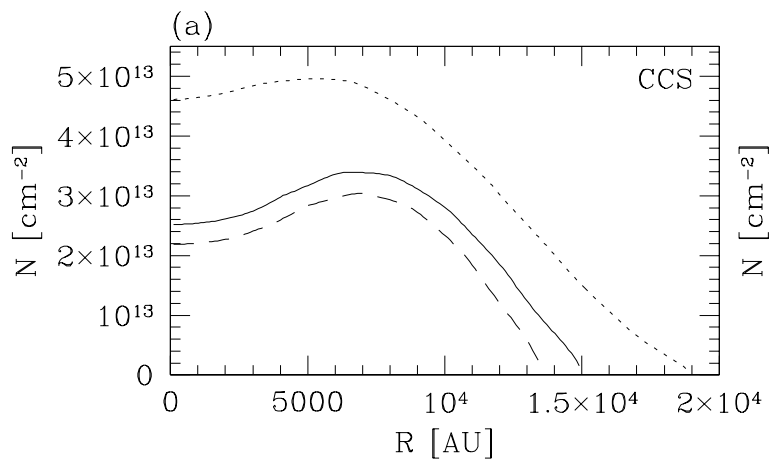
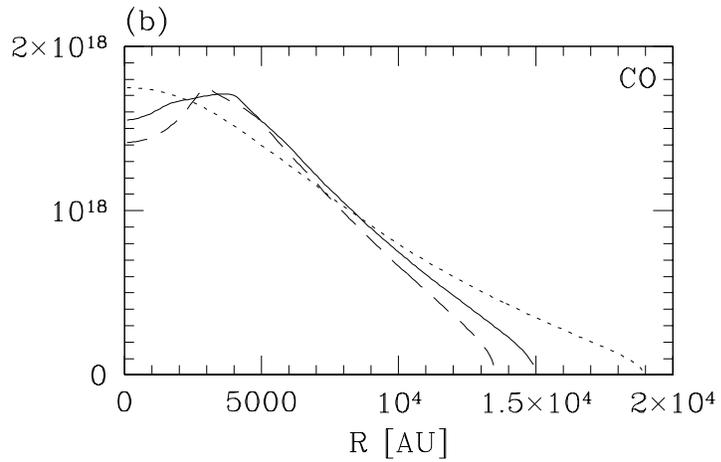
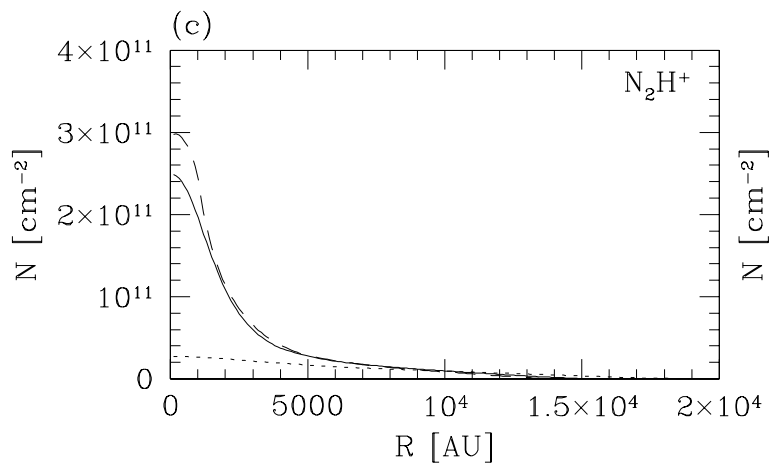
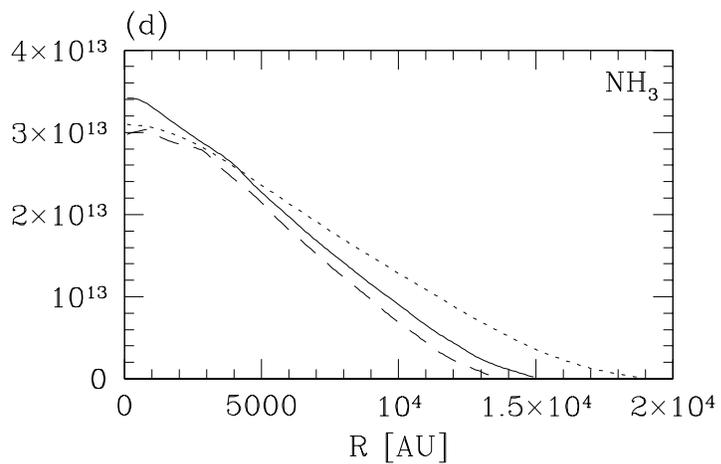
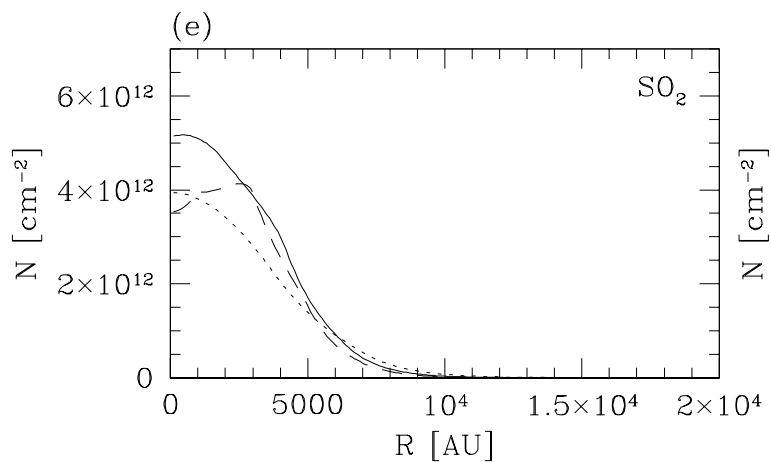
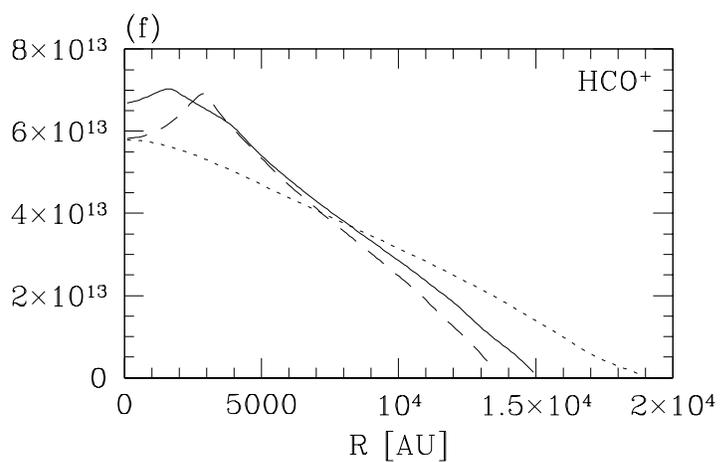
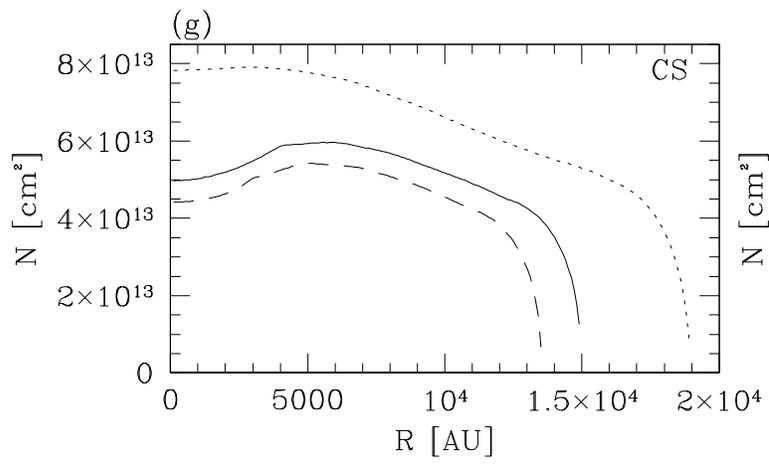
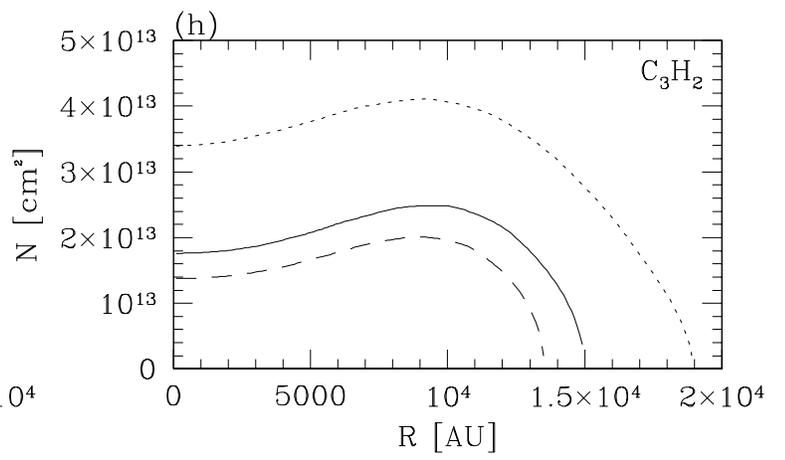
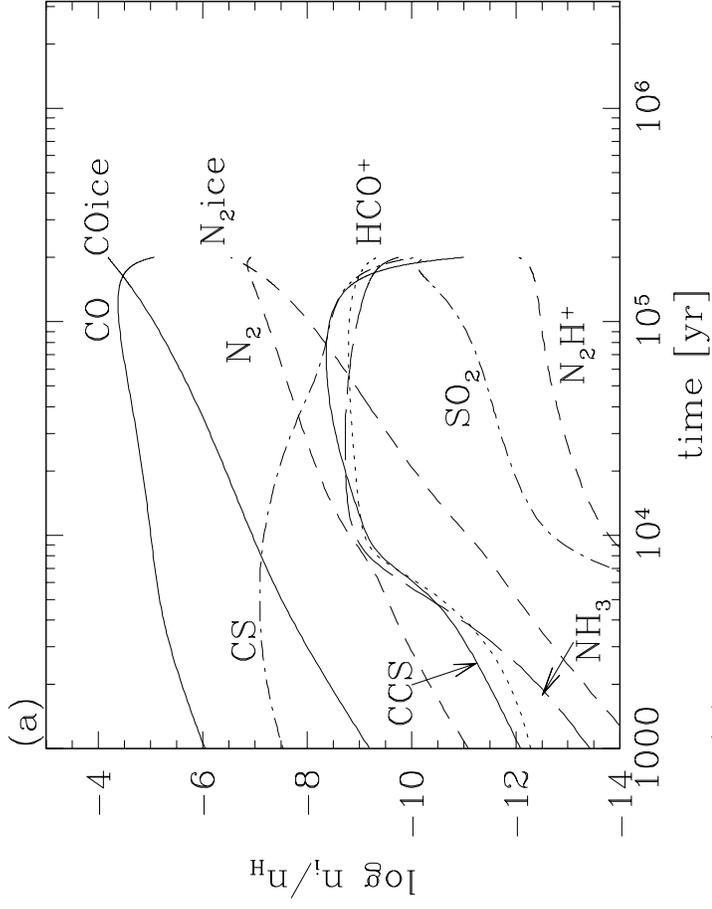
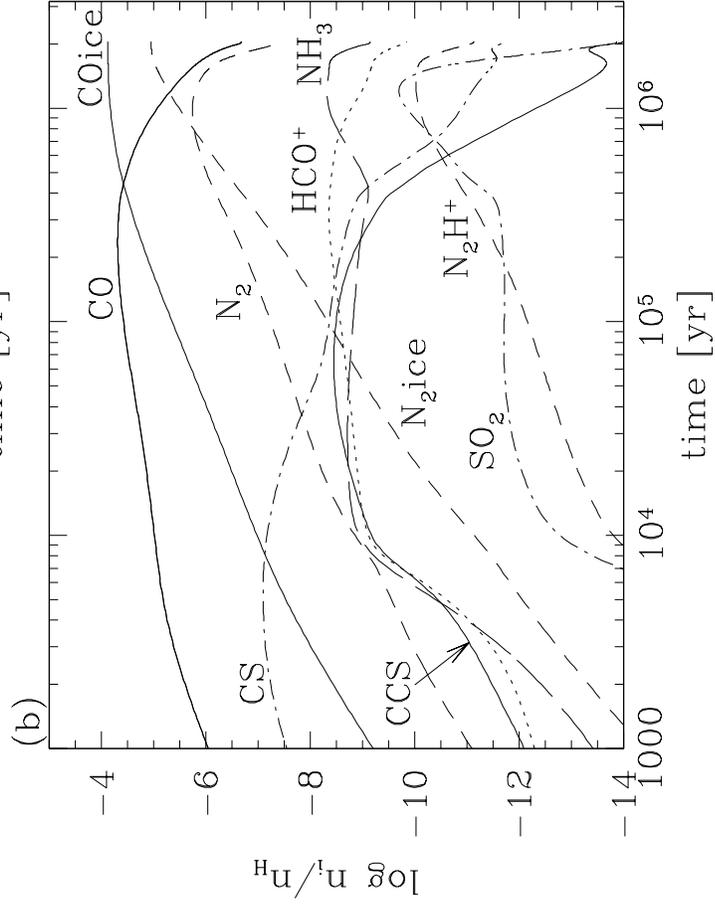
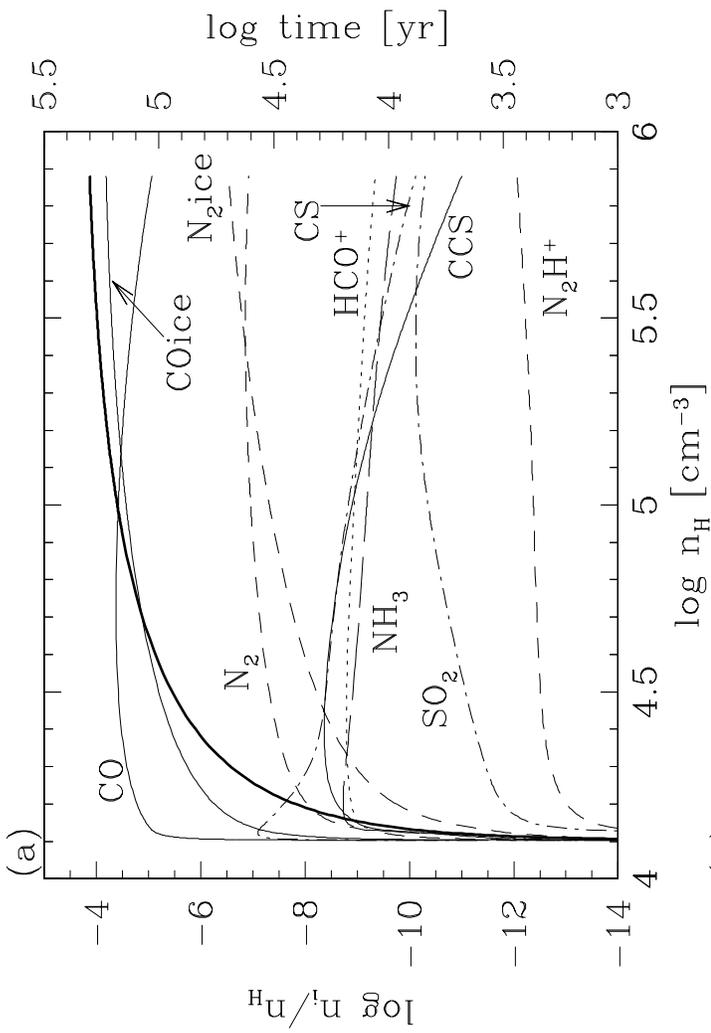
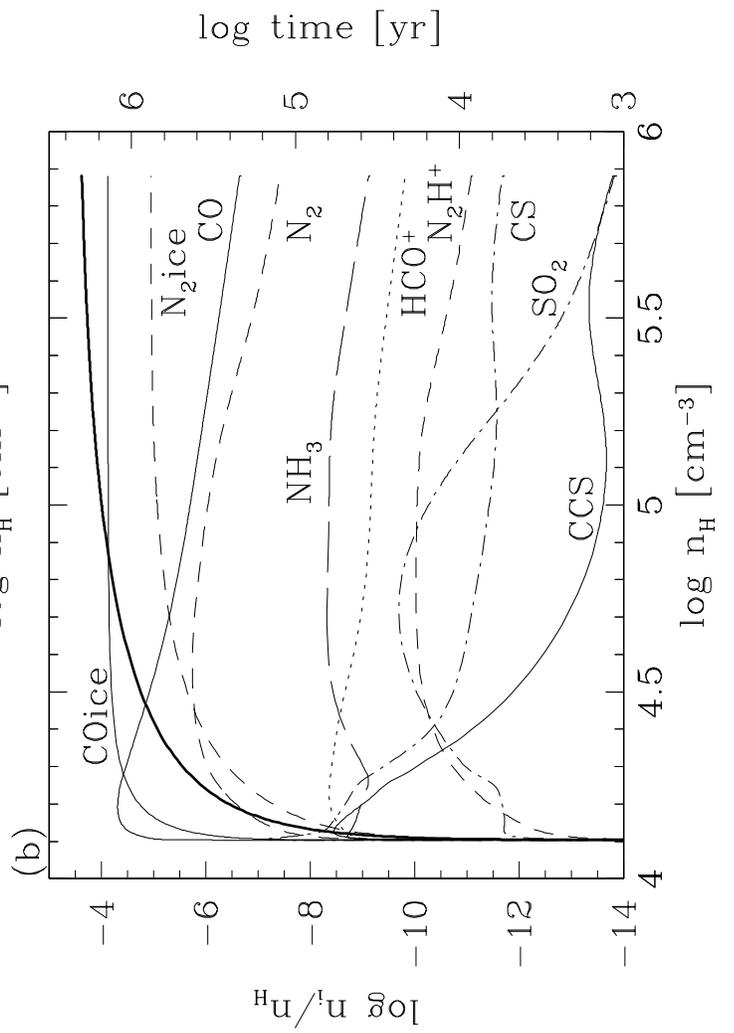
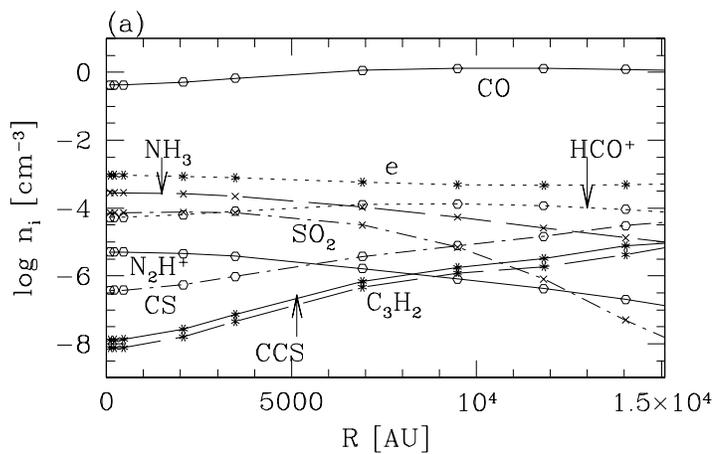
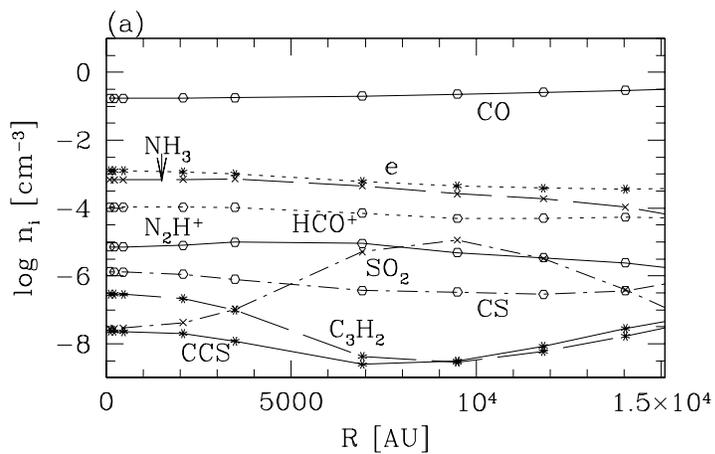
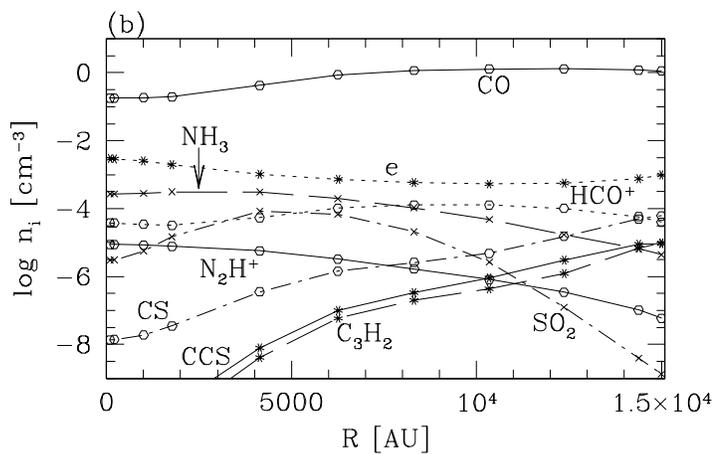
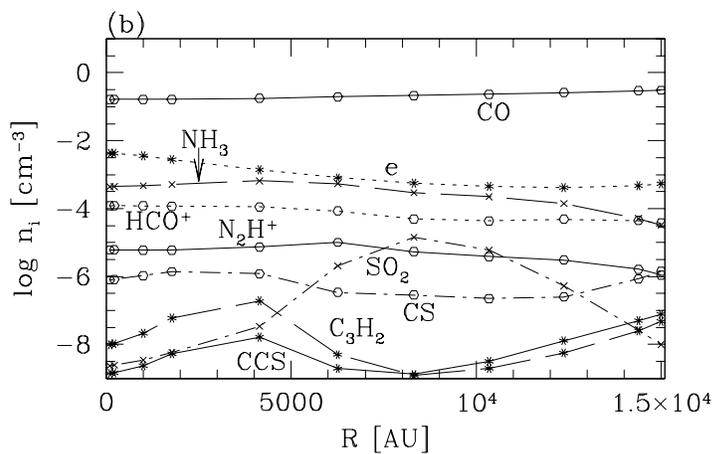
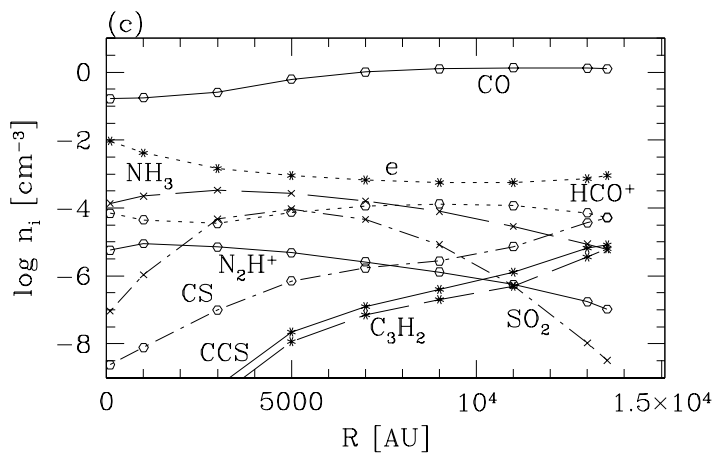
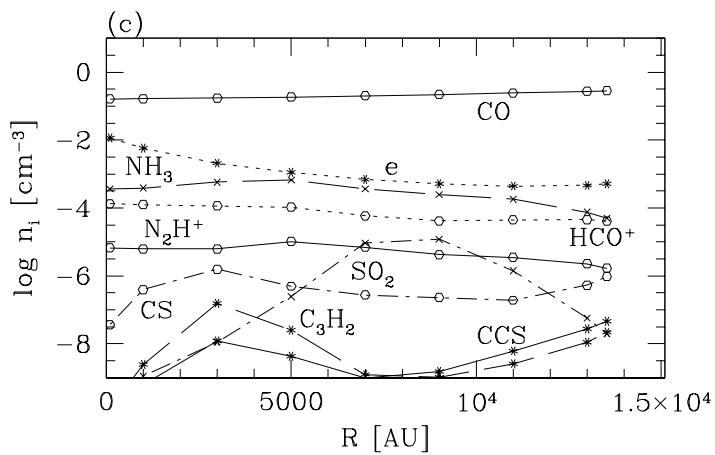
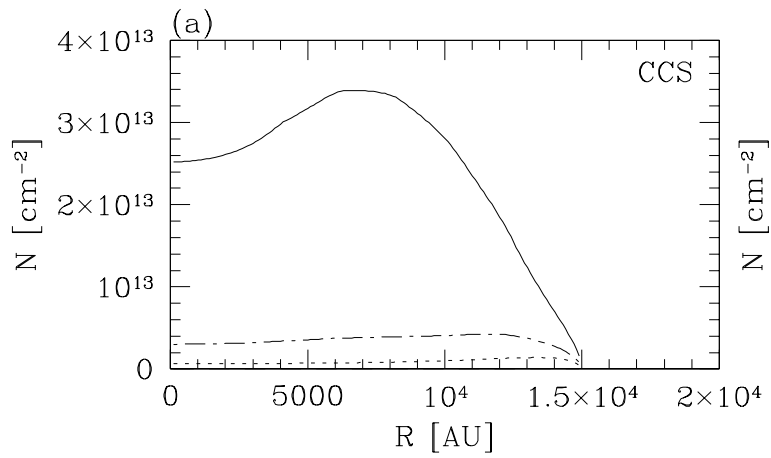
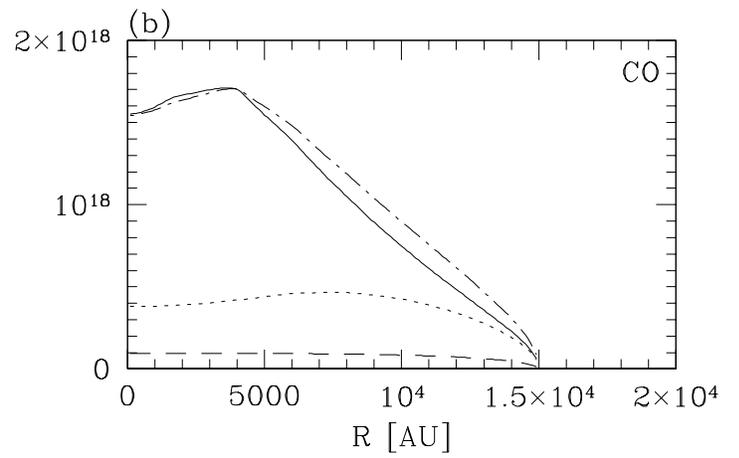
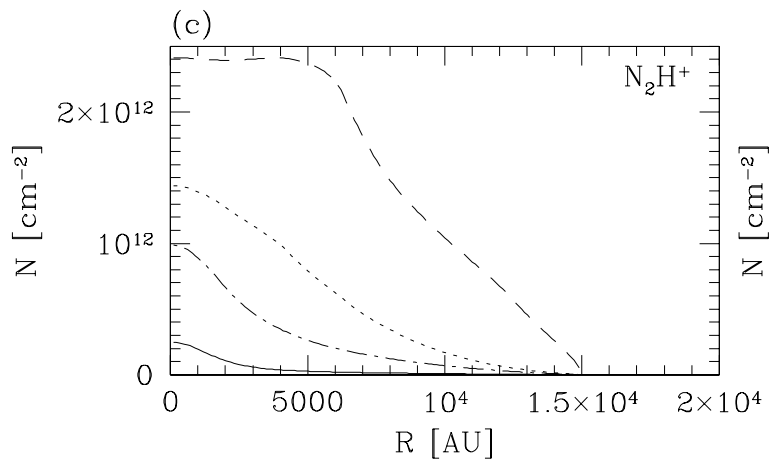
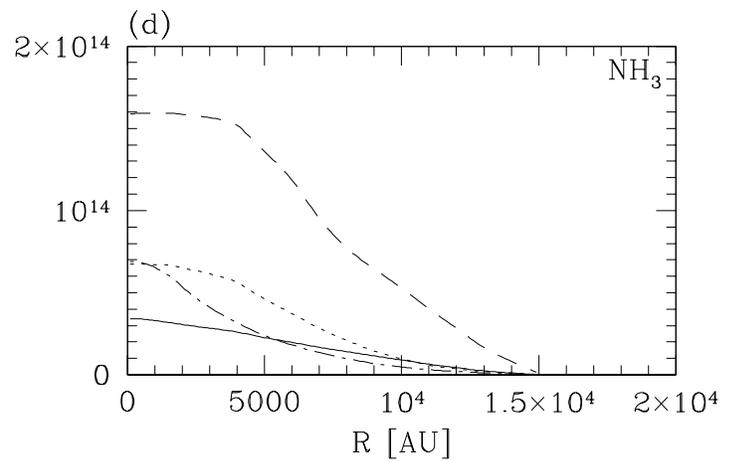
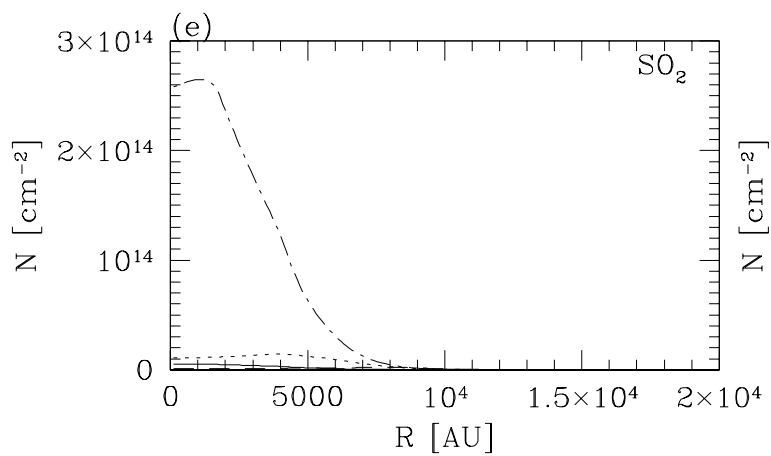
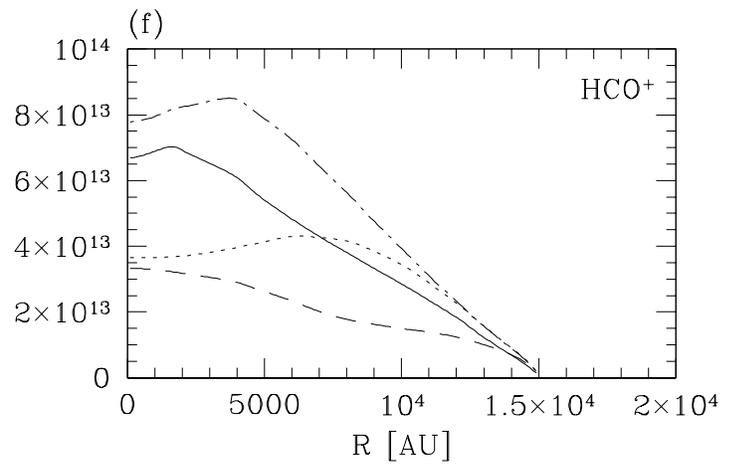
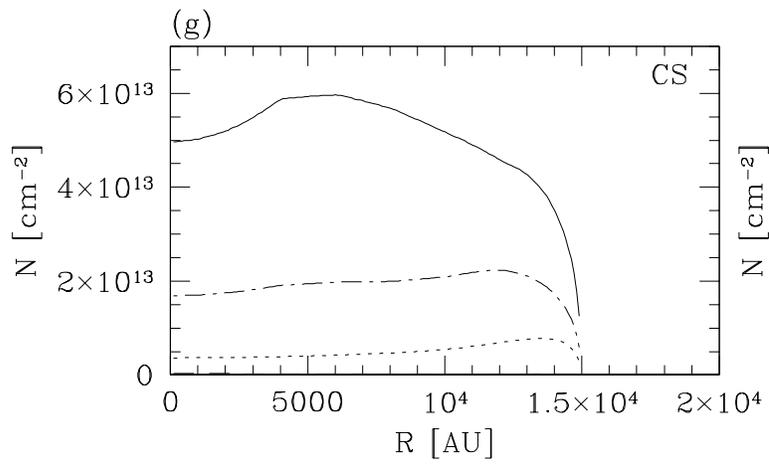
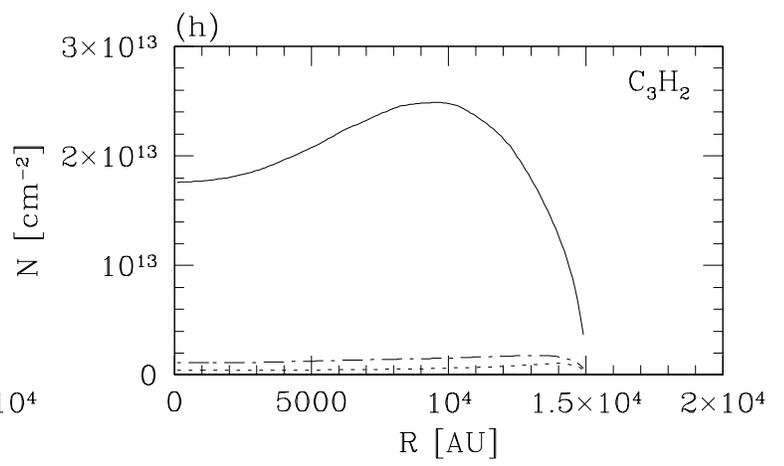
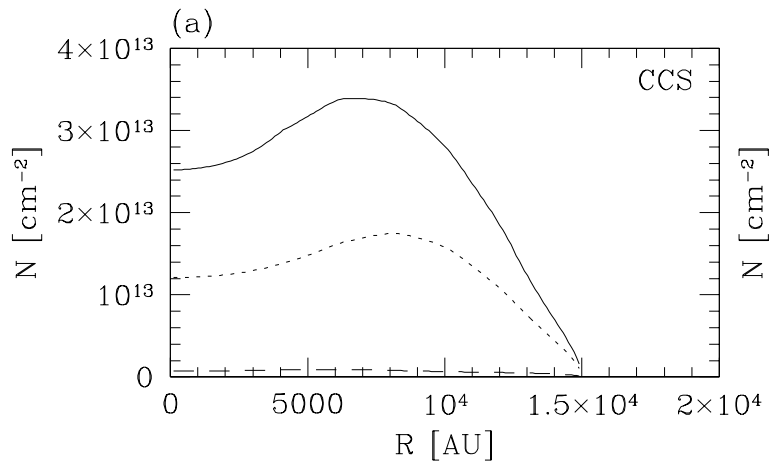
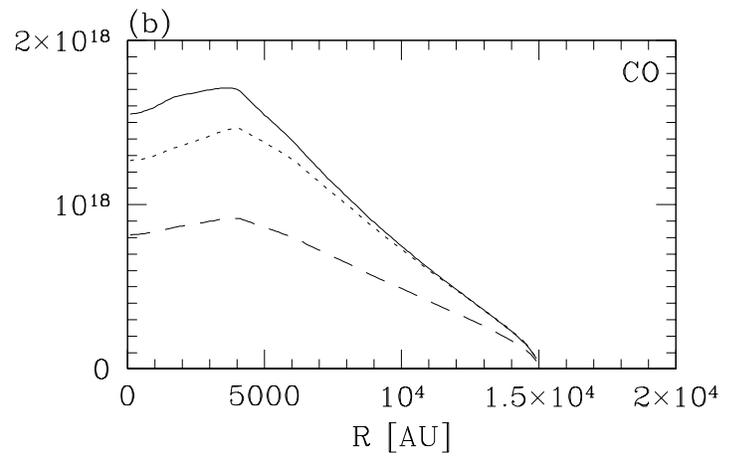
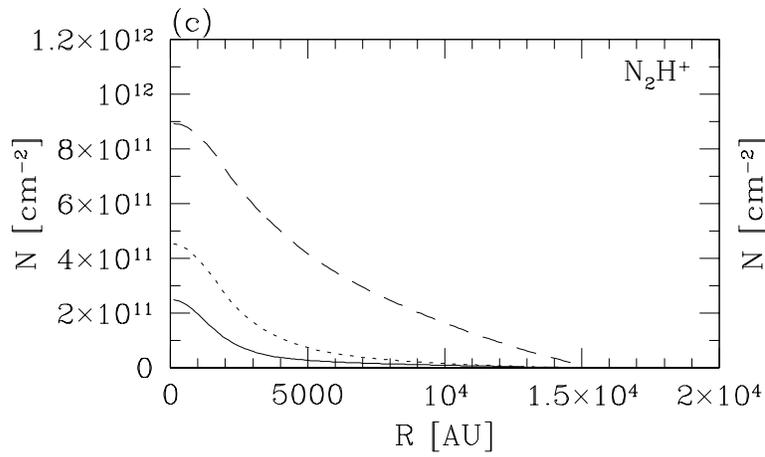
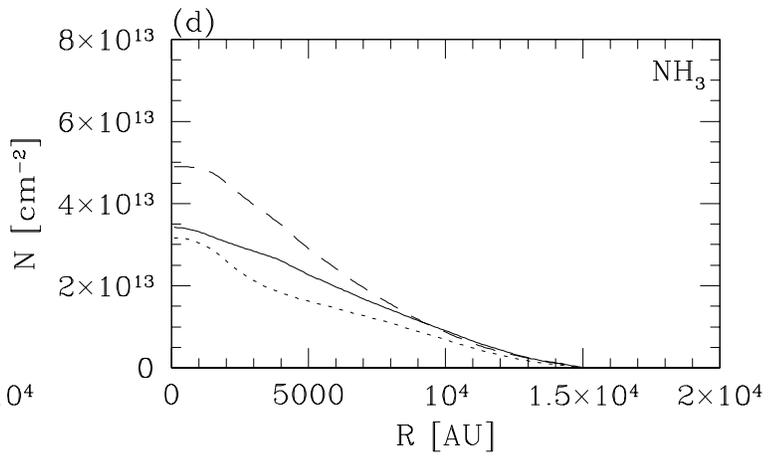
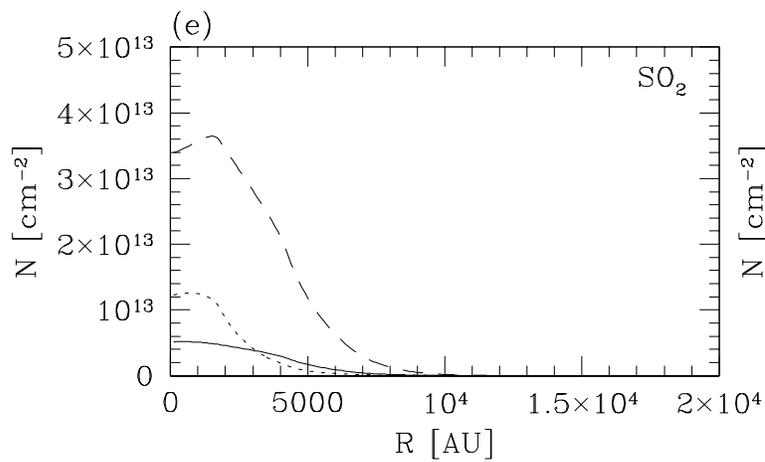
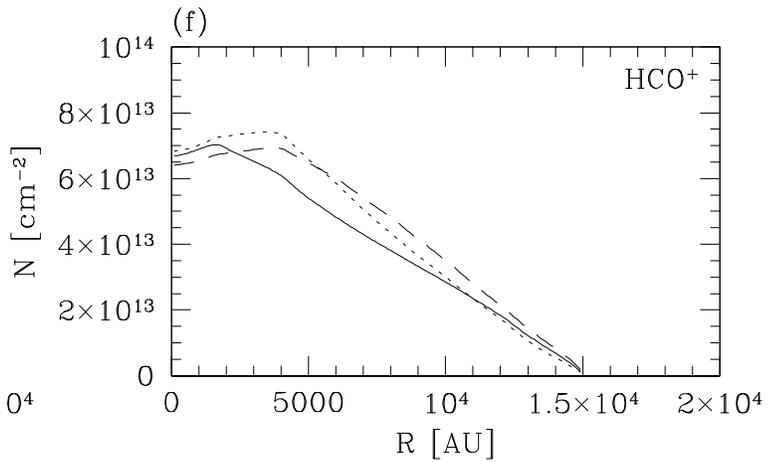
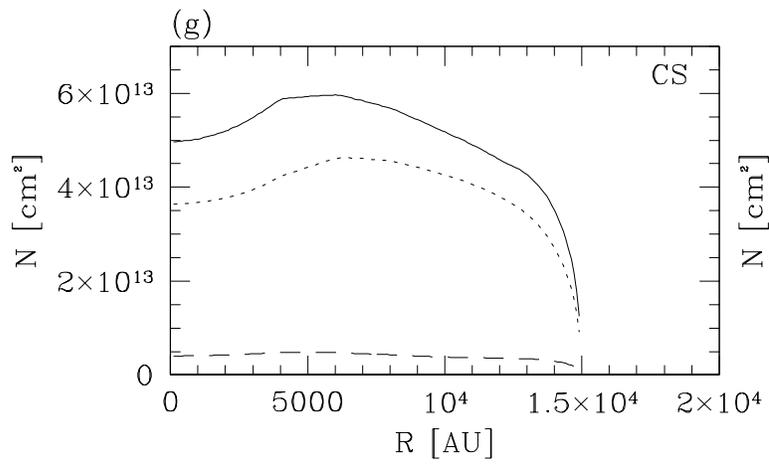
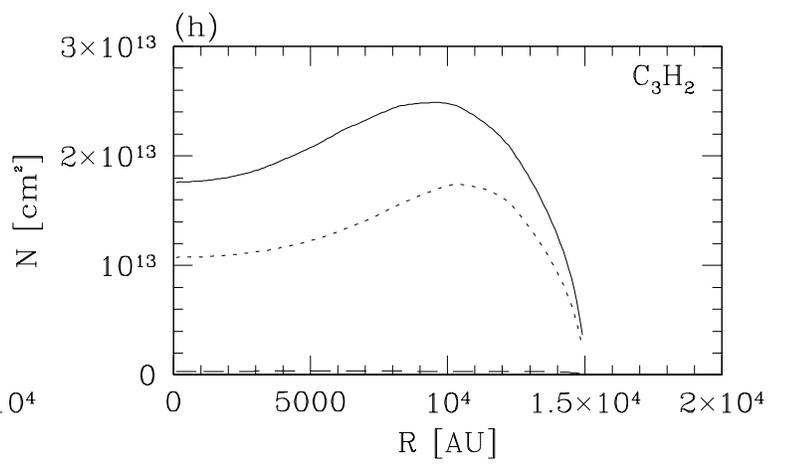
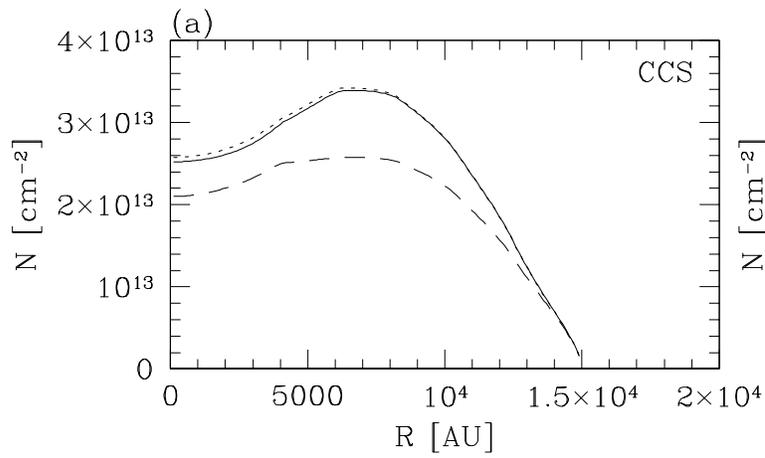
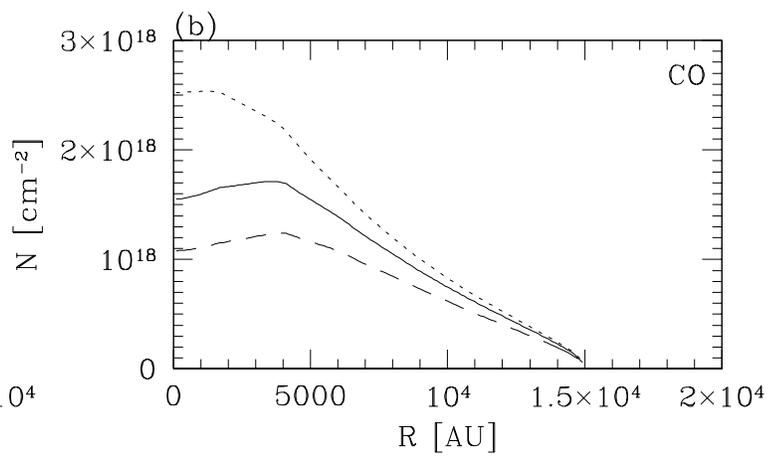
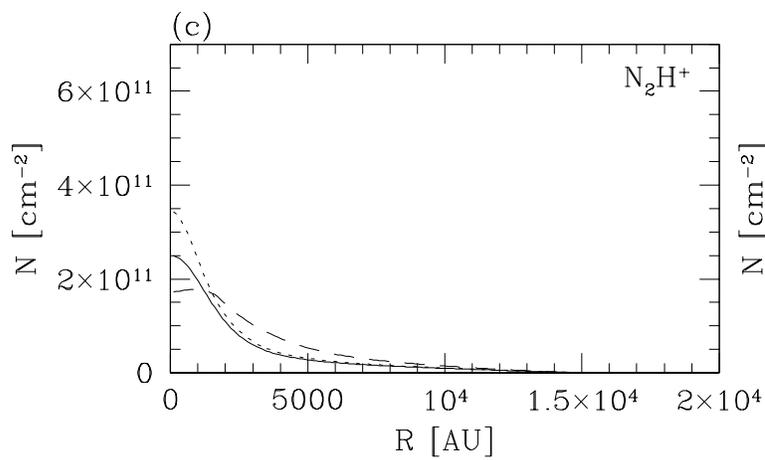
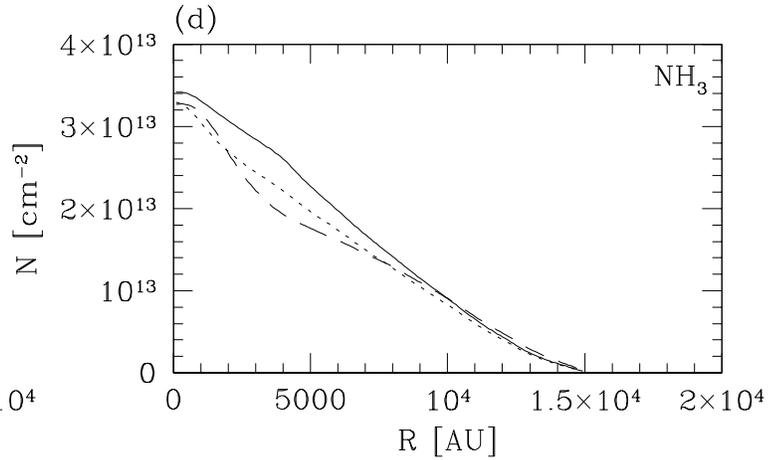
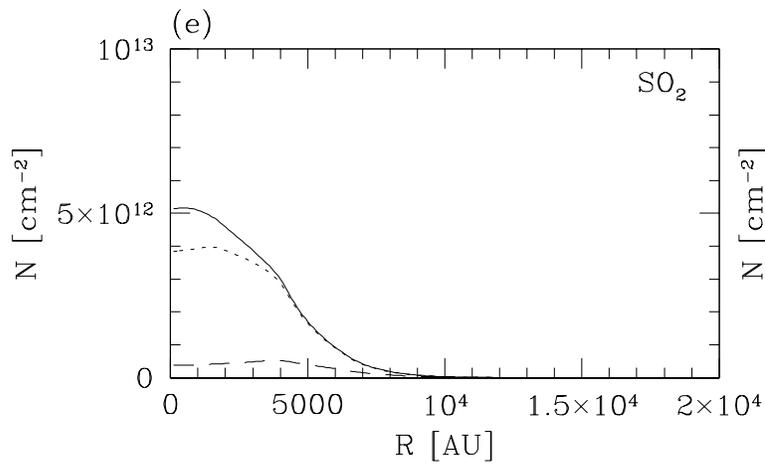
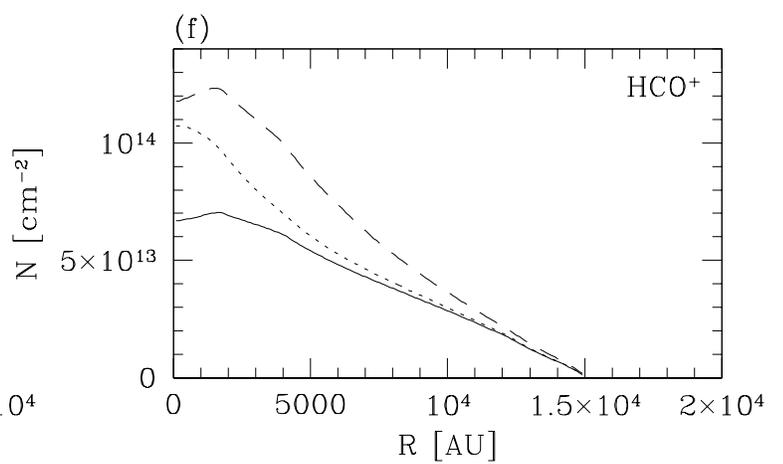
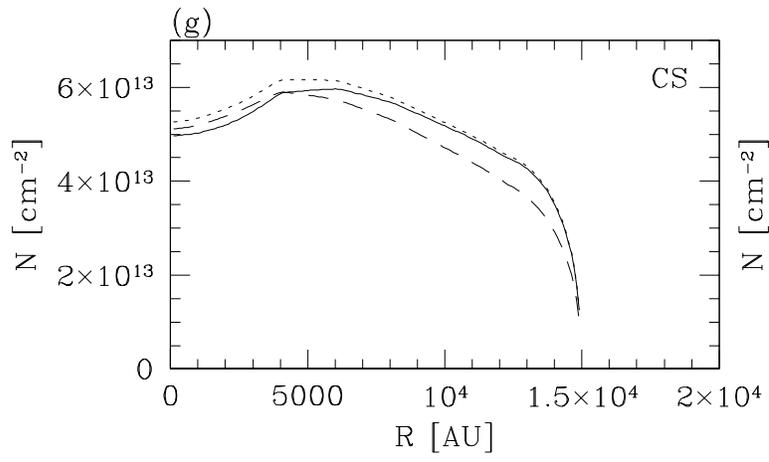
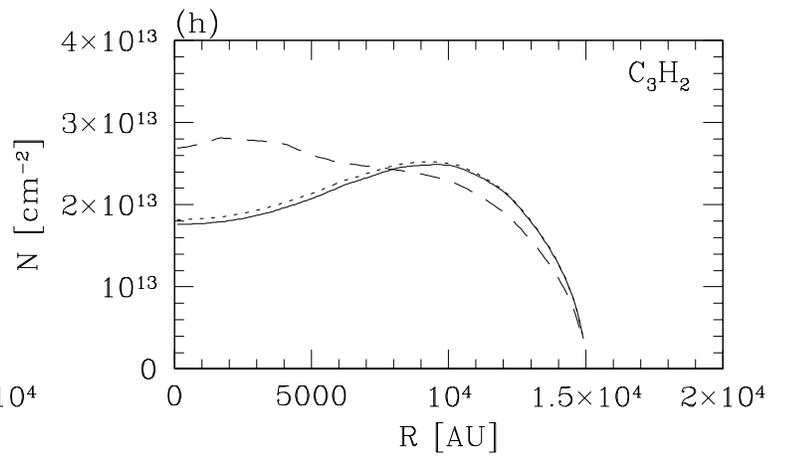
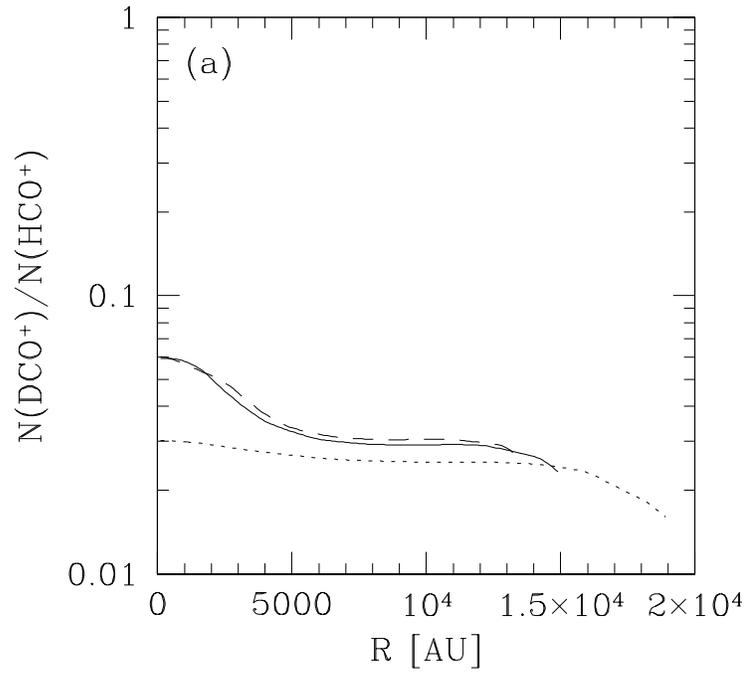
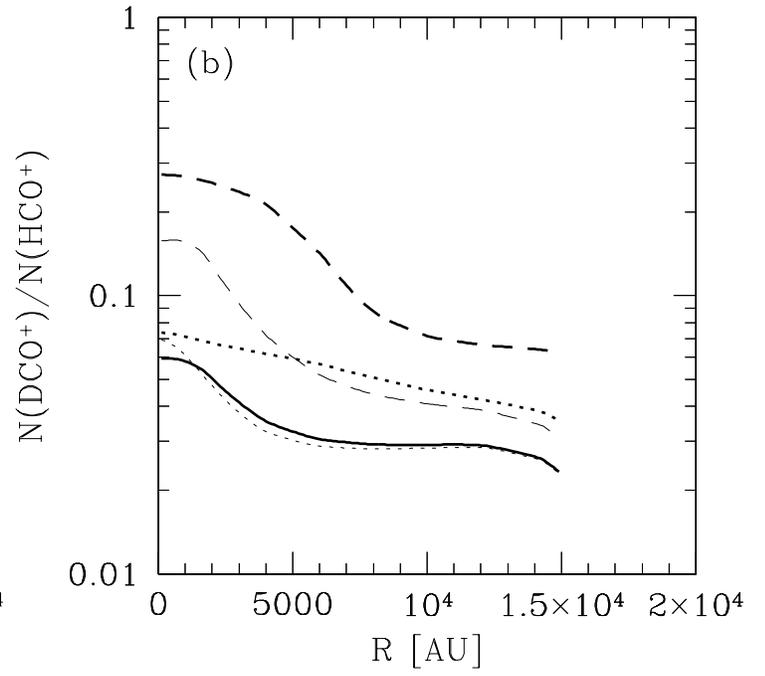
\end{document}